# Unravelling the dust attenuation scaling relations and their evolution

Gabriel Maheson ,[1,2]★ Roberto Maiolino,[1,2,3] Mirko Curti ,[1,2,4] Ryan Sanders ,[5] Sandro Tacchella [1,2] and Lester Sandles[1,2]

[1]Kavli Institute for Cosmology, University of Cambridge, Madingley Road, Cambridge CB3 0HA, UK
[2]Cavendish Laboratory, University of Cambridge, 19 J. J. Thomson Ave., Cambridge CB3 0HE, UK
[3]Department of Physics and Astronomy, University College London, Gower Street, London WC1E 6BT, UK
[4]European Southern Observatory, Karl-Schwarzschild-Strasse 2, D-85748 Garching, Germany
[5]Department of Physics and Astronomy, University of California, Davis, One Shields Ave, Davis, CA 95616, USA

## ABSTRACT

We explore the dependence of dust attenuation, as traced by the $H_\alpha/H_\beta$ Balmer decrement, on galactic properties by using a large sample of Sloan Digital Sky Survey spectra. We use both partial correlation coefficients and random forest analysis to distinguish those galactic parameters that directly and primarily drive dust attenuation in galaxies, from parameters that are only indirectly correlated through secondary dependencies. We find that, once galactic inclination is controlled for, dust attenuation depends primarily on stellar mass, followed by metallicity and velocity dispersion. Once the dependence on these quantities is taken into account, there is no dependence on the star formation rate. While the dependence on stellar mass and metallicity was expected based on simple analytical equations for the interstellar medium, the dependence on velocity dispersion was not predicted, and we discuss possible scenarios to explain it. We identify a projection of this multidimensional parameters space which minimizes the dispersion in terms of the Balmer decrement and which encapsulates the primary and secondary dependences of the Balmer decrement into a single parameter defined as the reduced mass $\mu = \log M_* + 3.67[O/H] + 2.96\log(\sigma_v/100\,km\,s^{-1})$. We show that the dependence of the Balmer decrement on this single parameter also holds at high redshift, suggesting that the processes regulating dust production and distribution do not change significantly through cosmic epochs at least out to $z \sim 2$.

## 1 INTRODUCTION

Although it only constitutes a small fraction of the mass in galaxies, interstellar dust plays an important role in the thermodynamics and chemistry of the interstellar medium, which impacts how galaxies form stars, as well as the reprocessing of light, shaping the galaxies spectral energy distributions (SEDs) (e.g. Conroy 2013). Dust can form in the atmospheres around AGB stars, or even in the ejecta of supernovae, and are then released into the interstellar medium (ISM) (Draine 2011). The formed dust grains have a range of sizes, typically from 5 to 250 nm (Weingartner & Draine 2001), and the dust grain size and mass evolve through various different mechanisms (Asano et al. 2013) such as growth by accretion from the ISM, coagulation, and destruction by shocks. Dust is formed of various chemical elements, such as Si, C, Fe, and $H_2O$ (Draine 2003), and so a fraction of the metals in galaxies is stored in the dust. Additionally, dust acts as a catalyst in the formation of $H_2$ regions, where the Hydrogen atoms adsorb onto the surface of the dust and chemically bond to form molecular Hydrogen (Draine 2003). An important cooling mechanism in galaxies is via dust (Montier & Giard 2004; Vogelsberger et al. 2019) through thermal infrared emission of heated dust grains. Cooling by dust facilitates gravitational collapse and fragmentation, hence the formation of (low mass) stars (Schneider et al. 2006).

The most important aspect of dust concerned in this work is its ability to scatter and absorb ultraviolet (UV) and optical light emitted by stars and by the ISM, and re-emit them at longer IR wavelengths, effectively reshaping the whole SED of the galaxy. These processes cause the effects known as dust extinction, dust attenuation and dust reddening (Draine 2003). Being able to accurately reverse the effects of dust on galaxy SEDs is essential to accurately determining galaxy parameters and investigating the physics within galaxies that drives their evolution.

Additionally, understanding how dust content scales with other galactic parameters can tell us a lot about how dust forms, and what role it plays in the evolution of galaxies. There are several scaling laws in the literature, which relate different galactic parameters, several of which relate (or may relate) to the dust mass, which in turn is related to the dust attenuation in the galaxies.

The star-forming main sequence (SFMS, Brinchmann et al. 2004; Sandles et al. 2022) relates the stellar mass to the star formation rate (SFR), and recent studies suggest that this is actually an indirect relation, a by-product of other more fundamental relations (Lin et al. 2019; Baker et al. 2022b on resolved scales, and Baker et al. 2023 on integrated scales). The relation between the SFR and the stellar mass is also present on resolved kpc to sub-kpc scales, forming the so-called resolved star-forming main sequence

★ E-mail: gm695@cam.ac.uk

(rSFMS, Sánchez et al. 2013; Cano-Díaz et al. 2016; Hsieh et al. 2017).

The resolved molecular gas main sequence (MGMS) relates the stellar mass surface density to the molecular gas mass surface density, as found in Lin et al. (2019), Barrera-Ballesteros et al. (2020), Morselli et al. (2020), Pessa et al. (2021), Ellison et al. (2021a, b), and Baker et al. (2022b). The molecular gas mass surface density is also related to the SFR surface density, and this is known as the Schmidt–Kennicutt law (SK, Schmidt 1959; Kennicutt 1998), which is understood as the molecular gas acting as fuel for star formation (Kennicutt 1998).

The fundamental metallicity relation (FMR) empirically relates the stellar mass, SFR and metallicity of the ISM, and has been reported by several authors (Mannucci et al. 2010; Nakajima & Ouchi 2014; Salim et al. 2014; Gebhardt et al. 2016; Hunt et al. 2016; Hirschauer et al. 2018; Curti et al. 2020a; Baker et al. 2022a), with the metallicity depending primarily on the stellar mass and showing a secondary, inverse dependence on the SFR. This is a generalization of the mass–metallicity relation (MZR, Lequeux et al. 1979) which is the correlation between the stellar mass and the metallicity. The FMR indicates a non-linear relationship between the stellar mass, metallicity and SFR, with the metallicity decreasing with SFR for low masses and becoming almost independent of SFR at higher masses. This relation is believed to hold true up to a redshift of $z \sim$ 3 (Cresci, Mannucci & Curti 2019; Sanders et al. 2021).

Various works have investigated the dependence of the dust attenuation, traced by the Balmer decrement ($H_\alpha/H_\beta$), on galactic properties. In particular, Garn & Best (2010) studied a local sample of galaxies at $z \sim 0.7$, investigating how the dust attenuation determined using the Balmer decrement method depends on the stellar mass, SFR and gas-phase metallicity. They determine a positive, non-linear correlation between the dust attenuation and each of these quantities. However, to understand which parameter is most important and driving the dust attenuation, and which of the other parameters only contribute a secondary dependence through the dependence on the other dominant parameter, if there is one, they employ principal component analysis (PCA). This method identifies which parameter causes the most variation in the dust attenuation. They determine the most important parameter to be the stellar mass, and they claim that the dependence of the dust attenuation on the other parameters are all secondary due to their dependence on stellar mass. They argue that the galaxies with a larger stellar mass will have built up a larger reservoir of dust, since dust is produced in stars (Draine 2003). However, the PCA method is only accurate when there is a simple linear relationship between the quantities, which is not necessarily present here.

Several other works have identified correlations between the dust attenuation and different galactic parameters, such as SFR (Garn et al. 2010), stellar mass (Pannella et al. 2009), and metallicity (Asari et al. 2007); however, few have investigated if the correlation identified is a direct correlation or if it is an indirect correlation introduced by secondary correlations between these and other galactic parameters.

To investigate how the dependencies between the dust attenuation and the galactic parameters evolve with redshift, giving insight into both the dust production mechanisms and how these evolve, works such as Shapley et al. (2022) have compared samples of local galaxies and higher redshift galaxies around cosmic noon. Cosmic noon, at $z \sim 2$–3, is the period where the cosmic average SFR was largest, and about 13 per cent of the stellar mass content of today's galaxies was formed (McLeod et al. 2021), making this epoch interesting for examining star formation mechanisms.

Several studies have identified the dependence of the dust attenuation on the stellar mass as the most important, and some have even observed this relation not evolving up to a redshift of about $z \sim 2$ (Whitaker et al. 2017; McLure et al. 2018; Shapley et al. 2022), and more recently up to $z \sim 6.5$ (Shapley et al. 2023). In particular, Shapley et al. (2022) identified no significant evolution in the relationship between the dust attenuation (using both the Balmer decrement and the UV continuum) and the stellar mass between Sloan Digital Sky Survey (SDSS) ($z \sim 0$) and MOSFIRE Deep Evolution Field (MOSDEF) ($z \sim 2.3$) galaxies, and argue that this lack of evolution can be explained by considering the evolution of the other parameters, such as metallicity, dust mass and gas mass.

There is some evidence that at $z > 3$ the dust attenuation evolves to lower values for a given stellar mass (Fudamoto et al. 2020). Following this, Bogdanoska & Burgarella (2020) compared samples of galaxies from the literature in the redshift range $0 < z < 10$ to identify how the UV dust attenuation versus stellar mass relation may evolve. They assumed a linear fit to the relation for all the samples and then tracked the evolution of this gradient with redshift. From this, they conclude that the relation between the dust attenuation and stellar mass evolves across the entire redshift range investigated, with the gradient of the linear fit peaking at cosmic noon and decreasing at higher and lower redshifts. This is in contrast to the results from Shapley et al. (2022), however, this work considered different samples of galaxies.

In this work, we wanted to first investigate how the Balmer decrement depends on the different galactic parameters, and in contrast to previous works, disentangle the various dependencies from primary and secondary relations using advanced statistical techniques, such as random forest (RF) regression and partial correlation coefficient (PCC) analysis. This part of the work was applied to large sample of local galaxies observed using the SDSS (York et al. 2000).

We also wanted to see if these relations evolve with redshift by comparing the local galaxies with samples of galaxies at higher redshift. In this work, we used two higher redshift samples, one observed by the *K*-band Multi Object Spectrograph (KMOS, Sharples et al. 2013) on the VLT, and the other observed by the Multi-Object Spectrometer for Infrared Exploration (MOSFIRE, McLean et al. 2012) on the Keck I telescope.

The layout of this paper is as follows. In Section 2, we discuss our data sources. In Section 3, we present the physics behind the determination of the dust attenuation from the emission-line fluxes. In Section 4, we make theoretical predictions on the most important galactic properties in determining the dust attenuation through known scaling laws. In Section 5, we present the statistical analysis tools used in this work. In Section 6, we present results of our statistical analysis and comparison between the samples of galaxies. In Section 7, we conclude the main findings of this work.

## 2 DATA AND SAMPLE SELECTION

In this work, we consider data from three spectroscopic surveys: SDSS, i.e. local galaxies at $z \sim 0$, and KMOS Lensed Velocity and Emission Line Review (KLEVER) and MOSDEF, i.e. galaxies at $z \sim 1$–3.

### 2.1 Local sample

To understand the dependency of the Balmer decrement on galaxy properties we first explored local galaxies, using the Sloan Digital

Sky Survey (SDSS, York et al. 2000) Data release 12 (DR12, Alam et al. 2015). SDSS uses a 2.5-m wide-field telescope at the Apache Point Observatory in New Mexico, US, utilizing the *u*, *g*, *r*, *i*, and *z* bands (Fukugita et al. 1996). The spectra of the objects are obtained by a pair of multiobject double spectrographs with 3-arcsec diameter fibres, producing a spectral coverage of 3800–9200 Å (Abazajian et al. 2009).

The spectroscopic redshifts, emission-line fluxes, stellar masses, and SFRs of the SDSS galaxies are calculated by the MPA-JHU group[1] providing measurements for 927 552 galaxies at redshifts $z < 0.7$.

### 2.1.1 Emission-line fluxes and nebular velocity dispersion

To determine the emission-line fluxes of the galaxies in the SDSS survey, the MPA-JHU group subtracted the best-fitting stellar population model of the stellar continuum from the spectra for each galaxy, then fit the nebular emission lines (Tremonti et al. 2004). To better measure the weak nebular lines, they fit Gaussians to the spectra simultaneously whilst requiring all the Balmer lines have the same width and velocity offset, and similarly for the forbidden lines, whilst also taking into account the spectral resolution. The nebular velocity dispersion was calculated from the spectra using the width of the emission lines.

### 2.1.2 Stellar mass

The stellar masses are taken from the MPA-JHU catalogue, and are calculated by comparing the observed SEDs of the galaxies to a large number of model SEDs, which then provides information on the stellar population of the galaxies, which are used to calculate the mass-to-light ratio, and hence the stellar mass of the galaxy (Kauffmann et al. 2003a; Salim et al. 2007) assuming the Kroupa (2001) initial mass function (IMF). These stellar masses were then converted to the Chabrier (2003) IMF.

### 2.1.3 Star formation rate

The SFR of galaxies can be calculated in several ways, one of which uses the H$_\alpha$ luminosity. Young massive stars, such as O/B, emit largely in the UV ionizing the gas around them, producing H II regions where recombination produces Balmer lines, with the H$_\alpha$ line being the brightest. Assuming all H$_\alpha$ emission is produced in H II regions around O/B stars and all the photons emitted by the star ionise the surrounding Hydrogen, the SFR can be shown to be proportional to the dust corrected H$_\alpha$ luminosity (Kennicutt & Evans 2012). The SFR data taken from the MPA-JHU catalogue was calculated using this method and assuming the Kroupa (2001) IMF, as is described in Brinchmann et al. (2004), with the H$_\alpha$ flux being dust corrected using the Balmer decrement method. These SFRs were then converted to the Chabrier (2003) IMF.

One issue of this method within the context of this paper is that the SFR derived from H$_\alpha$ will be strongly correlated with the Balmer decrement (H$_\alpha$/H$_\beta$), due to the H$_\alpha$ flux being dust corrected using the Balmer decrement, as well as the fact that H$_\alpha$ appears in both the SFR and the Balmer decrement. This cross-correlation between the SFR derived from H$_\alpha$ and the Balmer decrement is discussed in Appendix A. To avoid this spurious correlation affecting our results (especially with the Balmer decrement, which includes H$_\alpha$), another tracer of the SFR was used in this work, the SFR derived from fitting SEDs to the combined SDSS and WISE photometry from Chang et al. (2015). They fit the combined photometry using the SED-fitting code {\sc MAGPHYS} (da Cunha et al. 2012) to obtain monochromatic mid-IR SFR tracers, which do not suffer the same spurious correlations present in the SFR traced by the H$_\alpha$ flux.

**Table 1.** Table showing definitions of the emission-line ratios used to determine the metallicity of the galaxies through strong line calibrations. [O II]λλ3727,29 notation is equivalent to writing [O II]λ3727 + [O II]λ3729.

| Notation | Line ratio |
|---|---|
| R$_2$ | [O II]λ3727, 29/H$_\beta$ |
| R$_3$ | [O III]λ5007/H$_\beta$ |
| N$_2$ | [N II]λ6584/H$_\alpha$ |
| S$_2$ | [S II]λ6717, 31/H$_\alpha$ |
| R$_{23}$ | ([O II]λ3727 29 + [O III]λ4959, 5007)/H$_\beta$ |
| O$_3$O$_2$ | [OIII]λ5007/[O II]λ3727, 29 |
| RS$_{32}$ | [O III]λ5007/H$_\beta$ + [S II]λ6717, 31/H$_\alpha$ |
| O$_3$S$_2$ | [O III]λ5007/H$_\beta$/[S II]λ6717, 31/H$_\alpha$ |
| O$_3$N$_2$ | [O III]λ5007/H$_\beta$/[N II]λ6584/H$_\alpha$ |

We calculated an additional SFR tracer using the D4000 break following the methodology in Bluck et al. (2020a) to further investigate the importance of SFR in determining the Balmer decrement. This method is used in works such as Brinchmann et al. (2004) and Bluck et al. (2020a) when measurements on H$_\alpha$ are not available. The produced calibration and resulting analysis are shown in Appendix B, which support the conclusions of the main text.

### 2.1.4 Metallicity

The gas-phase metallicity of the galaxies in the SDSS survey was calculated for this work through the strong-line calibration method presented in Curti et al. (2020a). These calibrations between metallicity and ratios of strong emission lines were determined by using the 'direct' electron temperature ($T_e$) method, as reviewed in, for example, Maiolino & Mannucci (2019), where electron temperatures, $T_e$, are measured by stacking thousands of local galaxies to detect auroral lines and then used to infer the metallicity. The strong-line ratios used in this work are shown in Table 1, following the same definitions as are used in Curti et al. (2020a).

There are various caveats to the different diagnostics, such as some being double valued (Curti et al. 2020a). To mitigate these issues, we combine different combinations of the diagnostics for each galaxy as is done in Curti et al. (2020a). In this work, we chose to use the gas-phase metallicity relative to the solar metallicity, [O/H] = 12 + log(O/H) − 8.69 (Asplund et al. 2009).

### 2.1.5 Inclination

In this work, we explore the dependence of dust attenuation on galaxy inclination. The inclination measurements for the galaxies in this sample were extracted from the Simard et al. (2011) morphological catalogue. To determine the morphological parameters such as the galaxy inclinations, they used a galaxy model with the sum of a pure exponential disc and a de Vaucouleurs bulge (Sérsic index n$_b$ = 4).

### 2.1.6 Selection criteria

The DR7 sample consists of 927 552 galaxies. To reduce the effect of noise contributed by the sky background, the detector and the fluctuations in the source itself, we set signal-to-noise cuts on the line fluxes. Following Mannucci et al. (2010) and Hayden-Pawson et al. (2022), we adopt a high signal-to-noise ratio (S/N) on the $H_\alpha$ line ($>20\sigma$), which is essentially equivalent to a cut in SFR, and ensures that we have detections of $H_\beta$ and other metal lines needed to measure the metallicity without having to impose constraints on their S/N, which would bias the metallicity of the sample. We, however, do set a signal-to-noise cut on $H_\beta$ of $2\sigma$ to ensure the measured Balmer decrement is reliable. Imposing any higher signal-to-noise cuts on the $H_\beta$ line would bias our sample to less dusty galaxies, since dustier galaxies would attenuate the $H_\beta$ line more and so tend to have weaker $H_\beta$ detections. If the signal-to-noise cut on $H_\beta$ was increased to $3\sigma$, there would be 18 fewer galaxies in our sample, a 0.08 per cent decrease, which is such a small difference our results would not be affected by such a change in the selection criteria. In addition to these signal-to-noise cuts, when determining the metallicities of the galaxies, any diagnostic with lines detected above the three $\sigma$ level were combined to calculate the metallicity following Curti et al. (2020a).

The metallicity diagnostics used in this work relied on only star-forming galaxies (SFGs) in the sample, which were selected using BPT emission-line diagnostic diagrams (Baldwin, Phillips & Terlevich 1981), which compare the [O III]$\lambda$5007/$H_\beta$ line ratio against the [N II]$\lambda$6583/$H_\alpha$ line ratio. We did not apply signal-to-noise cuts on the galaxies for the [N II] and [O III] lines, since this would bias the metallicity measurements of the galaxies. The signal-to-noise cut of 20 on the $H_\alpha$ line implies the other metal lines used in Curti et al. (2020a) are detected. For the few cases, in which some of the metal lines are not detected, the upper limit on these lines can still be used to constrain the metallicity (Curti et al. 2020a). We used the Kauffmann et al. (2003b) demarcation line to define SFGs in this work. Due to the spectral resolution of SDSS being 2000, we also selected only galaxies with log nebular velocity dispersion above $\log_{10}(\sigma_{H_\alpha}[\mathrm{km/s}]) > 1.75$.

To ensure not just the central region of the galaxies was being sampled, we enforced the projected fibre aperture to be at least 2 kpc, which set a lower limit on the redshift of $z = 0.043$, since the aperture diameter was 3 arcsec. We set no upper cut on the redshift. As is explained in Section 6.1.1, the galaxies were also selected such that their inclination was less than 45°, to minimize the effect of increased dust attenuation with increased inclination. The result of these selection criteria reduced the local sample to 21 488 galaxies, with a maximum redshift of $z = 0.308$ and a median redshift of $z = 0.104$.

## 2.2 High-redshift samples

To see how the relations between dust attenuation and other galaxy properties evolve, we studied samples of galaxies at higher redshifts observed spectroscopically in the near-IR. The first sample we investigated at higher redshift was the near-infrared KMOS Lensed Velocity and Emission Line Review (KLEVER, Curti et al. 2020b). KLEVER is a European Southern Observatory Large Programme which has observed 192 galaxies in the redshift range $1.2 < z < 2.5$ using the KMOS) on the VLT (Sharples et al. 2013). KMOS is a near-IR multiobject spectrograph using integral-field units (IFUs) observing in the $Y$, $J$, $H$, and $K$ bands.

We additionally used galaxies taken from the MOSDEF (Kriek et al. 2015) survey to compare with our local galaxies (SDSS). This survey is observed with the Multi-Object Spectrometer for Infrared Exploration (MOSFIRE, McLean et al. 2012) on the Keck I telescope observing in the $Y$, $J$, $H$, and $K$ bands. The MOSDEF survey measured the rest-frame optical spectra of 1824 galaxies in the three redshift intervals, $1.37 < z < 1.7$, $2.09 < z < 2.61$, and $2.95 < z < 3.80$, which were selected such that the brightest rest-optical emission lines fall within the atmospheric transmission windows. Because of this, the galaxies in the higher redshift interval ($2.95 < z < 3.80$) were ignored in this analysis, as the $H_\alpha$ line was redshifted out of all the bands used on MOSFIRE. In this work, we only used the galaxies in the $2.09 < z < 2.61$ redshift bin.

The IFU observations from KMOS allowed us to perform very accurate measurements of the Balmer decrement at high redshift. MOSFIRE instead observes emission lines through single slits which require slit loss corrections, which are prone to vary with operation time. However, the effect of these slit loss corrections is shown to be insignificant in determining the Balmer decrement due to the relative flux calibrations between different bands agreeing to within 13 per cent by comparing the MOSFIRE spectra with photometric SED models (Kriek et al. 2015). The line flux measurements were additionally compared with 3D-HST grism line fluxes in Kriek et al. (2015) finding good agreement. The grism spectra do not have any slit aperture, much like IFU observations, which suggests the slit loss corrections are robust and the aperture affects are not significant.

Additionally, the galaxies in the KLEVER survey were selected using $H_\alpha$ emission in the rest-frame optical, from the 3D-HST survey (Curti et al. 2020b). The galaxies in the MOSDEF survey are instead selected using the flux in the $H$ band (dominated by the optical, rest-frame stellar continuum), aiming at obtaining a flat distribution in stellar mass (Kriek et al. 2015). The two selection criteria may potentially have different selection (bias) effects in terms of dust attenuation. The MOSDEF selection criteria do lead to a mass complete sample at $\log(M_\star/M_\odot) < 10.5$, however, at higher stellar masses the survey had its lowest mass completeness for red dusty SFGs (Kriek et al. 2015; Runco et al. 2022). Hence, the dust attenuation of the galaxies in the MOSDEF survey will be lower than expected at stellar masses above $10^{10.5}$ M$_\odot$. The MOSDEF sample has a lower mass completeness limit, however, with Sanders et al. (2021) not considering galaxies below $10^9$ M$_\odot$ due to them having a lower spectroscopic success rate when analysing the stellar MZR, and Shivaei et al. (2015) removing galaxies with stellar mass below $10^{9.5}$ M$_\odot$ due to poor photometric redshift measurements when analysing the SFR–stellar mass relation.

More recently, results have been reported for Balmer decrements at even higher redshift by using NIRSpec–JWST data (e.g. Shapley et al. 2023); however, not all the information required for our analysis (SFR, metallicity, and velocity dispersion) is provided for those galaxies, so they are not considered in this study. Moreover, the strongly wavelength-dependent slit losses of the small NIRSpec shutters, convolved with the galaxy sizes, make the uncertainties of line ratios spanning a large wavelength range uncertain. For these reasons, in this work, we mostly focus on the KLEVER and MOSDEF samples at $z \sim 1$–2.

In the following section, we provide additional information on how the measurements of the galaxy parameters used in this work were extracted from these two surveys.

### 2.2.1 KLEVER

The emission-line fluxes and widths were measured for the KLEVER survey as described in Hayden-Pawson et al. (2022). To determine the emission-line fluxes and widths, first a linear continuum was subtracted from the spectrum whose slope and normalization were free parameters. They did not fit a proper stellar continuum since the observed continuum was so faint. All emission lines within the same observing band were fit simultaneously with Gaussian curves of equal width, but across bands the widths were allowed to vary to account for different resolving powers within each band in KMOS. The nebular velocity dispersion was calculated in a consistent fashion with SDSS.

To correct for the stellar absorption of the Balmer emission lines (namely $H_\alpha$ and $H_\beta$), we calculated the stellar continuum from photometry for each galaxy. We used the Bayesian SED-fitting code {\sc BEAGLE} (Chevallard & Charlot 2016) to perform SED modelling of publicly available photometry (Merlin et al. 2016; Criscienzo et al. 2017; Bradač et al. 2019) for all of the objects in our sample with the aim of producing continuum-only spectra (without emission lines). We use a Chabrier (2003) IMF and assume a delayed exponential star formation history. Redshifts were fixed to their spectroscopic values. These continuum spectra were then normalized to the continuum around the Balmer lines in each band and subtracted from the integrated spectra. The $H_\alpha$, $H_\beta$, and [N II] doublet lines were simultaneously fit with Gaussians which had their redshift and widths tied. Three Gaussians were fit to deblend the $H_\alpha$ and [N II] doublet, and the amplitudes of the two [N II] Gaussians were fixed to have a ratio of 3:1.

The galaxy stellar mass is taken from the KMOS[3D] data release (Wisnioski et al. 2019) for KMOS, calculated through SED fitting following the methodology in Wuyts et al. (2011), similar to that in the SDSS sample. For the lensed galaxies in the sample, the stellar masses were calculated in Concas et al. (2022) by using SED fitting following the methodology in Curti et al. (2020b).

We calculated the metallicity measurements for KLEVER following the same choice of metallicity diagnostics as for SDSS Curti et al. (2020a).

The KLEVER sample provided by Hayden-Pawson et al. (2022) consisted of 192 galaxies. The only selection criteria we enforce are that both the $H_\alpha$ and $H_\beta$ lines are clearly detected. As a criterion, we conservatively use both the uncertainty from the Monte Carlo fitting (requiring an S/N > 2) and the nominal uncertainties on the spectrum, summed in quadrature over the FWHM of the line and centred at the nominal line location (requiring S/N > 3). The uncertainty from the Monte Carlo fitting of the fluxes is determined by perturbing the spectra with Gaussian noise randomly extracted from the noise spectra, repeating the fit one hundred times, and taking the 16th and 84th percentiles of the $2.5\sigma$ clipped resulting distribution of fluxes. In addition to this, when determining the metallicities of the galaxies, any diagnostic with lines detected above the $3\sigma$ level were combined to calculate the metallicity following Curti et al. (2020a). We then required the galaxies to have at least two diagnostics available, else the resulting metallicity may have strong biases.

The BPT selection that was applied to the galaxies in the SDSS survey is not valid at this redshift, since the demarcation derived by Kauffmann et al. (2003b) is for local galaxies. Measurements of galaxies around $z \sim 2$ in these BPT diagrams have identified an offset with the SDSS galaxies which can be mass dependent, for example, Shapley et al. (2015), and can be due to the higher redshift galaxies having a harder stellar ionizing radiation field and a higher ionization parameter (Steidel et al. 2014; see references therein). To remove AGNs and ensure only SFGs were considered in our analysis of the galaxies from the KLEVER survey, we first used the information from the [N II]–BPT diagram valid at the redshifts we are considering (Kewley et al. 2013), removing galaxies that were largely above the AGN demarcation line. We then implemented a visual inspection of all spectra to identify type 1 AGNs (via the presence of broad components under Balmer lines, mainly $H_\alpha$). We additionally cross-matched X-ray catalogues, classifying a galaxy as an AGN and removing from our sample if the X-ray luminosity was greater than $2 \times 10^{42}$ erg$^{-1}$. After these cuts, 51 galaxies were left in our sample.

### 2.2.2 MOSDEF

The emission-line fluxes were measured for the MOSDEF survey as described in Kriek et al. (2015). The systemic redshift was measured from the highest signal-to-noise emission line, usually the $H_\alpha$ or the [O III]λ5008 line. Line fluxes were measured by fitting Gaussian profiles over a linear continuum, where the centroids and widths were allowed to vary. Uncertainties on the line fluxes were estimated using a Monte Carlo method where the spectrum was perturbed according to the error spectrum and the line fluxes were remeasured. This process was repeated 1000 times, and the uncertainty on the line flux was taken to be the 84–16th percentile range of the resulting distribution. This method also produced the lines respective FWHMs, which were converted to the velocity dispersion. This data are publicly available.[2]

Since a stellar continuum was not able to be measured, the Balmer lines were corrected for underlying stellar atmospheric absorption (typically from type A stars) by modelling the galaxy stellar populations, as described in Reddy et al. (2015).

The stellar mass measurements for the MOSDEF galaxies were calculated by Sanders et al. (2021) using SED fitting to the photometry for the galaxies in the $z \sim 2.3$ redshift interval. This data was made available upon direct request to the authors.

The metallicity measurements for MOSDEF were calculated in this work following the same metallicity diagnostics as for SDSS and KLEVER from Curti et al. (2020a). The median of the metallicities $(12 + \log(O/H))$ for the galaxies selected from the MOSDEF survey calculated in this work is 8.43 with a 16–84th percentile range of 0.23. The metallicities calculated for the same sample of galaxies in Sanders et al. (2021) have a median of 8.47 with a 16–84th percentile range of 0.35. The metallicity measurements calculated in this work are slightly lower with a narrower spread than those from Sanders et al. (2021). However, to reduce biases in the choice of metallicity diagnostics used, since Sanders et al. (2021) calculated their own calibrations for their metallicity measurements, we chose to use the same method of determining metallicities for all galaxies used in this work, following Curti et al. (2020a).

AGNs were identified and removed from the galaxies used in this work provided by Sanders et al. (2021) using their X-ray and infrared properties, as well as their value of log ([N II]/$H_\alpha$) > −0.3 (Coil et al. 2015; Azadi et al. 2017).

The only flat S/N cuts were made on the $H_\alpha$ and $H_\beta$ lines, setting them to be greater than three for each line. In addition to this, when determining the metallicities of the galaxies following the same metallicity diagnostics as for SDSS and KLEVER from Curti et al. (2020a), any diagnostic with lines detected above the $3\sigma$ level were combined to calculate the metallicity, requiring the galaxies to have at least two diagnostics available. These cuts reduced the sample to 188 galaxies.

## 3 BALMER DECREMENT, REDDENING, AND DUST ATTENUATION

In this work, to measure the dust attenuation, $A_\lambda$, from observational data we use the Balmer decrement method. The Balmer decrement is defined as the ratio of the flux from the $H_\alpha$ to the $H_\beta$ emission lines. If we assume a Case B recombination, temperature of $T = 10^4$ K and an electron density of $n_e = 10^2 \mathrm{cm}^{-3}$, as is done in many similar studies (Garn & Best 2010; Piotrowska et al. 2020; Reddy et al. 2020), the Balmer decrement has an intrinsic value of 2.86. Recent work (Tacchella et al. 2022) suggests that the intrinsic Balmer decrement may be slightly higher when the contribution to the Balmer decrement from collisional ionization is taken into account.

---

[2]https://mosdef.astro.berkeley.edu/for-scientists/data-releases/

To determine the dust attenuation $A_\lambda$ from these Balmer-line fluxes, we first define the attenuation curve, $k_\lambda$, related to the dust attenuation and the reddening, $E(B - V)$, through the definition

$$k_\lambda = A_\lambda / E(B - V). \qquad (1)$$

Some attenuation laws (Calzetti et al. 2000; Reddy et al. 2015) are derived using empirical methods consisting in comparing the observed galaxy's SEDs with SEDs of galaxies which are assumed to be unattenuated. Another method to determine the attenuation law is SED fitting to a model of galaxy spectra built theoretically (Buat et al. 2012; Kriek & Conroy 2013). Both these methods are explained in depth in the review by Salim & Narayanan (2020).

It can be shown that the dust attenuation $A_\lambda$ is related to the Balmer decrement through

$$A_\lambda = -2.5 \frac{k_\lambda}{k_{\mathrm{H}_\alpha} - k_{\mathrm{H}_\beta}} \log_{10}\left( \frac{F_{\mathrm{H}_\alpha} / F_{\mathrm{H}_\beta}}{2.86} \right) \qquad (2)$$

where $k_{\mathrm{H}_\alpha}$ and $k_{\mathrm{H}_\beta}$ are the values of the attenuation curve at each Balmer wavelength, and $F_{\mathrm{H}_\alpha}$ and $F_{\mathrm{H}_\beta}$ are the observed flux of each Balmer line. Due to this relation, the Balmer decrement is a measure of the reddening of the spectra; to then measure the dust attenuation one must assume an attenuation curve, which itself could depend on attenuation (e.g. Chevallard et al. 2013; Tacchella et al. 2022). To reduce additional assumptions, in this work, we decided to not adopt a specific attenuation law and instead use the Balmer decrement itself to probe the dust attenuation and its dependence on the different galaxy parameters.

# 4 EXPECTED ATTENUATION DEPENDENCIES ON GALACTIC PROPERTIES

Considering the galactic scaling laws from the literature, it is possible to make theoretical expectations on how the dust content, hence the dust attenuation and its observational proxy, the Balmer decrement, scales with the galactic parameters discussed so far.

We expect that the dust attenuation $A_\lambda$, scales with dust mass, $M_d$, as well as geometric factors, $\gamma_g$. Hence, $A_\lambda \propto M_d \gamma_g$. Geometrical factors, such as the configuration of the stars, gas and dust within the galaxies, are difficult to constrain and are not considered in depth in this work. However, we can relate the dust mass to the other galactic properties. First, the dust mass scales with the mass of the gas in the galaxy, $M_g$, through the dust-to-gas ratio, DGR, allowing us to write $M_d = \mathrm{DGR} \times M_g$. Additionally, the dust mass scales with the mass of the metals, $M_Z$, through the dust-to-metal ratio, DZR, allowing us to write $M_d = \mathrm{DZR} \times M_Z$. We therefore can relate the DGR to the gas metallicity ($Z_g = \frac{M_Z}{M_g}$) and DZR, giving

$$\mathrm{DGR} = \frac{M_d}{M_g} = \frac{M_d}{M_Z} \times \frac{M_Z}{M_g} = \mathrm{DZR} \times Z_g. \qquad (3)$$

We can relate the gas mass to the stellar mass through the MGMS (Lin et al. 2019), $M_g = k_{\mathrm{MGMS}} \times M_\star$, where $k_{\mathrm{MGMS}}$ is a proportionality constant. Hence, we can write the following equation to determine how the dust mass, and hence dust attenuation, should scale with the

galactic parameters

$$M_d = \mathrm{DZR} \times Z_g \times k_{MGMS} \times M_\star \qquad (4)$$

where the metallicity $Z_g$ depends strongly on the stellar mass, and has a secondary inverse correlation with SFR (FMR, Curti et al. 2020a). Since the DZR is approximately constant these relations suggest the stellar mass will be the most important parameter in determining the dust mass, and so the dust attenuation. This follows from the assertion that all dust is produced in stars, and so if a galaxy has a larger stellar mass, it will likely have more dust. Equation (4) also suggests the dust attenuation depends directly on the gas metallicity, $Z_g$, and indirectly on SFR from the inversese secondary dependence of $Z_g$ on the SFR (FMR).

# 5 DATA ANALYSIS METHODS AND STATISTICAL ANALYSIS

Based on the simple modelling and assumptions described in Section 4, we have identified some of the galaxy properties which should be observationally more strongly related to dust content – this includes the stellar mass, SFR and metallicity, as is also suggested in Garn & Best (2010). To determine which are most important for our local sample of galaxies (SDSS), in this work, we combine PCC analysis and RF analysis. These two methods are described in the following sections.

## 5.1 PCC analysis

PCC analysis (Lawrance 1976) is a useful tool to describe the correlation between two quantities whilst controlling for the others. This allows us to disentangle primary correlations from indirect, secondary, correlations.

The PCC for variable A with variable B, fixing for variable C, $\rho_{\mathrm{AB|C}}$, is related to the Spearman rank correlation coefficient between A and B, $\rho_{\mathrm{AB}}$, and other combinations of the correlations between these variables. Specifically,

$$\rho_{\mathrm{AB|C}} = \frac{\rho_{\mathrm{AB}} - \rho_{\mathrm{AC}} \rho_{\mathrm{bc}}}{\sqrt{1 - \rho_{\mathrm{AC}}^2} \sqrt{1 - \rho_{\mathrm{bc}}^2}} \qquad (5)$$

as in Lawrance (1976).

We recall that the use of the Spearmann rank correlation is advantageous over Pearson correlation since the Spearmann rank correlaion first rank orders the parameters, which reduces the assumption of linearity between the parameters and instead favours monotonicity, which is useful in this work due to the non-linearity of many of the predicted relations (Bluck et al. 2020a; Baker et al. 2022b). See Baba, Shibata & Sibuya (2004) for further details.

The PCCs can be expanded to include more than three variables by using the methods provided in the *pingouin* (Vallat 2018) package. Yet, controlling for only the two most important variables is often adopted, as this maximizes performance and accuracy.

These coefficients can also be used to identify the direction of maximum variance of variable C in the parameter space defined by parameters A and B. On a plot of variable A on the $y$-axis against variable B on the $x$-axis, with variable C on the $z$-axis (e.g. colour-coded), an arrow can be drawn in the $x$–$y$ plane with angle $\theta$ clockwise from the positive $y$-axis, denoting the direction of maximum variation, or largest gradient, in variable C. Such arrow angles can be quantified by using the PCCs through the following equation:

$$\tan\theta = \frac{\rho_{\mathrm{AC|B}}}{\rho_{\mathrm{bc|A}}}, \qquad (6)$$

adapted from Piotrowska et al. (2020) and Bluck et al. (2020a).

To determine the error on the PCCs and $\theta$ in this method when applied to the galaxies in the samples, bootstrap random sampling was used, taking 100 random samples of the data with replacement, with each sample the same size as the original data set, and computing the standard deviation on these results, as is done in Baker et al. (2022b).

We note that PCCs can provide a useful indication of the direct correlations as long as these are monotonic.

## 5.2 RF analysis

In this work, we also use RF analysis. This is a widely used machine-learning method, which uses decision trees to determine which parameters are most important in predicting the target variable for a set of data. The decision trees work by trying to reduce the Gini Impurity at each branch of the decision trees (Pedregosa et al. 2011). The parameter importances can then be determined by averaging their contribution to the decrease in Gini Impurity from all of the decision trees in the forest, representing which parameters were used most to predict the value of the target variable.

We used RF regression to predict the parameter importances in determining the target variable (the Balmer decrement in this work). The data are split into a train and test sample, with a 50:50 split, where the train sample is used to train the regressor, and the test sample used to evaluate the accuracy of the regressor. We used the RF regressor from the {\sc PYTHON} package Scikit-learn (Pedregosa et al. 2011).

Compared to PCC analysis, RF analysis does not require the variables to have a monotonic relationship and can simultaneously explore the dependence on multiple intercorrelated quantities (Bluck et al. 2020a, b). PCC analysis additionally tells us the direction of the dependence, where RF does not.

To maximize the efficiency and accuracy of the regressor, we fine-tune specific execution parameters within the RF function, known as hyperparameters. In this work, we fine-tuned the hyperparameter dictating the minimum number of samples allowed to exist at the end of a decision tree, in other words, controlling how many splits the decision tree is allowed to make when training the regressor, which in turn controls for the size the decision tree is allowed to grow to. This hyperparameter is known as the minimum number of samples on the final leaf, where leaf refers to a final node of the decision tree. If this hyperparameter is set too low, the regressor has a tendency to overfit to the training sample by splitting at every opportunity until the training data is unphysically fit, however, if the value is too large the accuracy of the regressor is low, as it has not had enough time to fit to the training data. The results of the fine-tuning for the local galaxies are presented in Appendix C.

The error on the determined importances were calculated by repeating the whole process 100 times, re-splitting the data and re-training the regressor each time. The error on the importances was then taken as the standard deviation of the calculated importances for each parameter.

For further detail on RF analysis, see Bluck et al. (2022).

## 6 RESULTS

In this section, we present the results of our statistical analysis using both PCC and RF. We also explore the dependence of the Balmer decrement on the various parameters and identify projections of these multidimensional parameter spaces that minimize the scatter of the individual relations, hence finding analytical relations that simultaneously describe the dependence of the Balmer decrement on multiple galactic quantities.

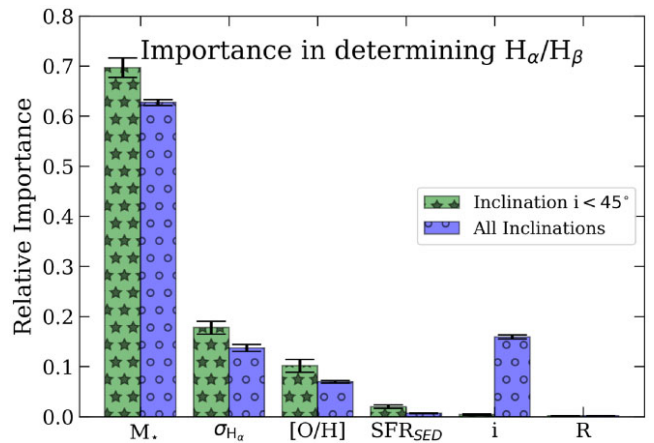

**Figure 1.** Relative importance of the different galactic parameters in determining the Balmer decrement for local galaxies (SDSS) calculated using RFs. Green bars with stars show importances for galaxies with inclination $i < 45°$, while blue bars with circles show the importance for galaxies with all inclinations. Stellar mass is the most important parameter in both samples. Selecting for the inclination $i < 45°$ reduces the importance of the inclination to be negligible. SFR$_{SED}$ is the SFR calculated using SED fitting. The parameter R is a control, random variable.

We then use these projections to compare the derived analytical relations between local galaxies (SDSS) and galaxies at $z \sim 1$–$3$ (KLEVER and MOSDEF).

### 6.1 SDSS

#### 6.1.1 Random forest

We investigate the importance of various galactic parameters in predicting the Balmer decrement using RF regressors in order to explore the theoretical framework described above. Specifically, we investigate the importance of the following galactic parameters: stellar mass ($M_*$), velocity dispersion inferred from H$\alpha$ ($\sigma_{H\alpha}$), gas-phase metallicity ([O/H]), SFR calculated using SED fitting (SFR$_{SED}$), inclination of the galaxy (i) and a control random variable (R), calculated using the {\sc NUMPY} (Harris et al. 2020) random number generator uniformly between 0 and 1, to test how valid the calculated importances are. The details of the tuning of the hyperparameters used in the RF, in order to increase the accuracy of the regressor, can be found in Appendix C.

The importance of the velocity dispersion of the stars is additionally investigated in Appendix D, showing it has a similar importance compared to the nebular velocity dispersion, however, stellar velocity dispersion data were not available for the higher redshift galaxies, and so was not used further in this work.

We do not include the SFR inferred from the H$\alpha$ emission line as the same quantity enters into the Balmer decrement. Together also with the fact that the H$\alpha$ flux is further corrected for extinction through the Balmer decrement, these aspects result in a spurious correlation with the Balmer Decrement itself. This is, however, discussed in Appendix A.

The RF regressor was first run on our sample of local galaxies with no pre-selection on their inclination. The importance of each parameter, along with its error, in determining the Balmer decrement is shown in Fig. 1 as the blue bars with circles. This tells us that the stellar mass is the most important parameter in determining the Balmer decrement by far (consistent with previous studies e.g. Garn & Best 2010), followed by the inclination, $i$.

From Fig. 1, we can see that the inclination is very important. This makes sense physically, since if we see a galaxy edge-on, the flux from the galaxy will encounter more dust on average as it travels to us compared with if the galaxy was face-on, implying the dust attenuation will be larger. The effect of inclination on the observed galaxy properties has been studied in depth for the galaxies in the SDSS survey, for example, Maller et al. (2009), which investigated the correlation between inclination and both the colour and size of the galaxies, in order to determine inclination corrections to these parameters. They show that the correction in the $g$ band can reach up to 1.2 mag, with the corrections not depending much on the galaxy luminosity, but depending strongly on the Sérsic index. Hence, this importance is not intrinsic to how the dust attenuation or dust content is related to the galaxy properties, but simply consequence of the viewing angle. To reduce this effect and focus on more fundamental parameters, we pre-select galaxies in terms of their inclination such that its importance was as small as possible whilst maintaining a large enough sample of galaxies to which we can apply our statistical tools to confidently. We investigated this by cutting the inclination to be less than 60°, 45°, and 30°, with 90° being edge-on and 0° being face-on. The inclination importance dropped significantly, and we found a sample with inclination less than 45° had inclination with negligible importance (as quantified further below) whilst maintaining a large sample. This cut the sample from 65 613 to 21 488 galaxies, whilst maintaining all other signal-to-noise selection criteria discussed in Section 2.1.6.

Using the sample of local galaxies with inclination $i < 45°$ we recalculated the importance of each galactic parameter in determining the Balmer decrement. These importances and their errors are shown as the green bars with stars in Fig. 1. The stellar mass is still the most important parameter, now followed by the velocity dispersion and then by the metallicity. The SFR has very little importance, barely above the random variable, $R$. The inclination is now as important as the random variable, $R$. Therefore, this selection on the inclination was adopted for this work, and henceforth all local galaxies analysed will have inclinations $i < 45°$ unless specified otherwise, to control for its effect on the Balmer decrement.

### 6.1.2 2D histogram visualization and PCC arrows

To better visualize the relative importance of the parameters, in Fig. 2, we plot the local galaxies used in this work (with inclination $i < 45°$) in a hexagonal 2D binning scheme, since hexagons allow for better data aggregation around the bin centre than rectangular bins. Since the stellar mass is found by the RF to be the most important parameter, we keep the $x$-axis as stellar mass, and vary the $y$-axis between the SFR$_{SED}$, metallicity and the nebular velocity dispersion. The dependent variable (i.e. the one for which we want to find the dependence on the other quantities), on the $z$-axis (colour-coded), is always the Balmer decrement. The galaxies were binned in hexagonal bins, and the median Balmer decrement of the galaxies in each bin was calculated. Bins with less than 25 galaxies were ignored. The contours show the density of the galaxies in this space, with the outermost contour containing 95 per cent of the galaxies in the sample. The contours in Fig. 2(c) do not connect due to the sharp cut in the velocity dispersion, producing a discontinuity in the density distribution.

The PCC arrows were calculated using the binned galaxy parameters rather than the individual galaxies, to avoid the analysis being dominated by the inner, most populated regions. Here, the PCC-derived arrows indicate the direction in which the Balmer decrement has the largest average gradient on the 3D surface of each diagram, with its angle defined clockwise from the positive $y$-axis. The error on the angles was calculated through bootstrap random sampling.

The colour-shading and gradient arrows visually illustrate how the Balmer decrement depends on all of these parameters with varying strength. Considering the angles of the arrows on each of the plots, the Balmer decrement has a strong correlation with the stellar mass. The colour shading and PCC arrow in panel (a) visually confirms that, at fixed stellar mass, there is essentially no dependence of the Balmer decrement on SFR. Panels (b) and (c) visually show that, at a fixed stellar mass, the Balmer decrement also depend significantly both on the metallicity and on the velocity dispersion. The inclination of the PCC arrow in (b) and (c) being close to 45° would naively indicate that the dependence on metallicity and velocity dispersion is stronger than inferred from the RF, and nearly as strong as the dependence on the Stellar Mass. However, one has to take into account that these 2D histograms only consider the dependence on only two quantities at a time, therefore any residual dependence not associated with the quantities on the plot must be taken by one of them. Therefore, it is likely that in plot (b) the metallicity it also picking the Balmer decrement dependence on the velocity dispersion, while vice versa in plot (c) the velocity dispersion is picking the Balmer decrement dependence on the metallicity, if metallicity and velocity dispersion are correlated with each other.

### 6.1.3 Partial correlation coefficients

In this section, we further investigate the importance of the various parameters identified in the previous sections by using the PCC analysis on all parameters whilst keeping the two most important parameters constant. With respect to the RF analysis, the (full) PCC additionally tells us the direction (sign) of the dependence. As before, we apply this analysis to both samples of local galaxies with and without a selection on their inclination in order to explore its effect.

To determine the PCCs for this sample of local galaxies with no cut on their inclination, we kept the two most important parameters constant. The PCCs between the Balmer decrement and the galaxy parameters deemed important in this analysis (stellar mass, metallicity, velocity dispersion, inclination, and the SFR calculated using SED fitting) are shown in Fig. 3. Similar analysis including the SFR derived from $H_\alpha$ is shown in Appendix A, supporting the conclusion that the relation between the Balmer decrement and SFR$_{H_\alpha}$ is driven mostly by the fact the Balmer decrement was used to dust correct the $H_\alpha$ flux in SFR$_{H_\alpha}$, as well as $H_\alpha$ also appearing in the Balmer decrement. The green bars with stars show the PCCs using galaxies with inclination $i < 45°$, and the blue bars with circles represent the PCCs using the galaxies with no selection on their inclination.

For the sample with no selection on its inclination, the strongest correlation with the Balmer decrement is with the stellar mass and then the inclination, which is consistent with the RF results shown in Fig. 1. The PCCs for the galaxies with inclination $i < 45°$ show that the stellar mass is still the most strongly correlated parameter with the Balmer decrement, now followed by the metallicity and the velocity dispersion, with the inclination being much less correlated compared with the sample with no selection on its inclination. Hence, these results also support the selection on the inclination of the local galaxies, allowing the effects of the inclination on the Balmer decrement to be controlled for. The exact values of the PCC and the RF importances for each galaxy parameter may vary between the methods and depend on which sample is being analysed. This is due to the PCC analysis only controlling for the two most important parameters in each sample and so the exact value of the PCCs will still be affected by secondary correlations (which are therefore picked by the two most important parameters), which instead the RF analysis is able to take into account. Hence, minor differences in the RF importances and PCC values for the galaxy parameters are to be expected.

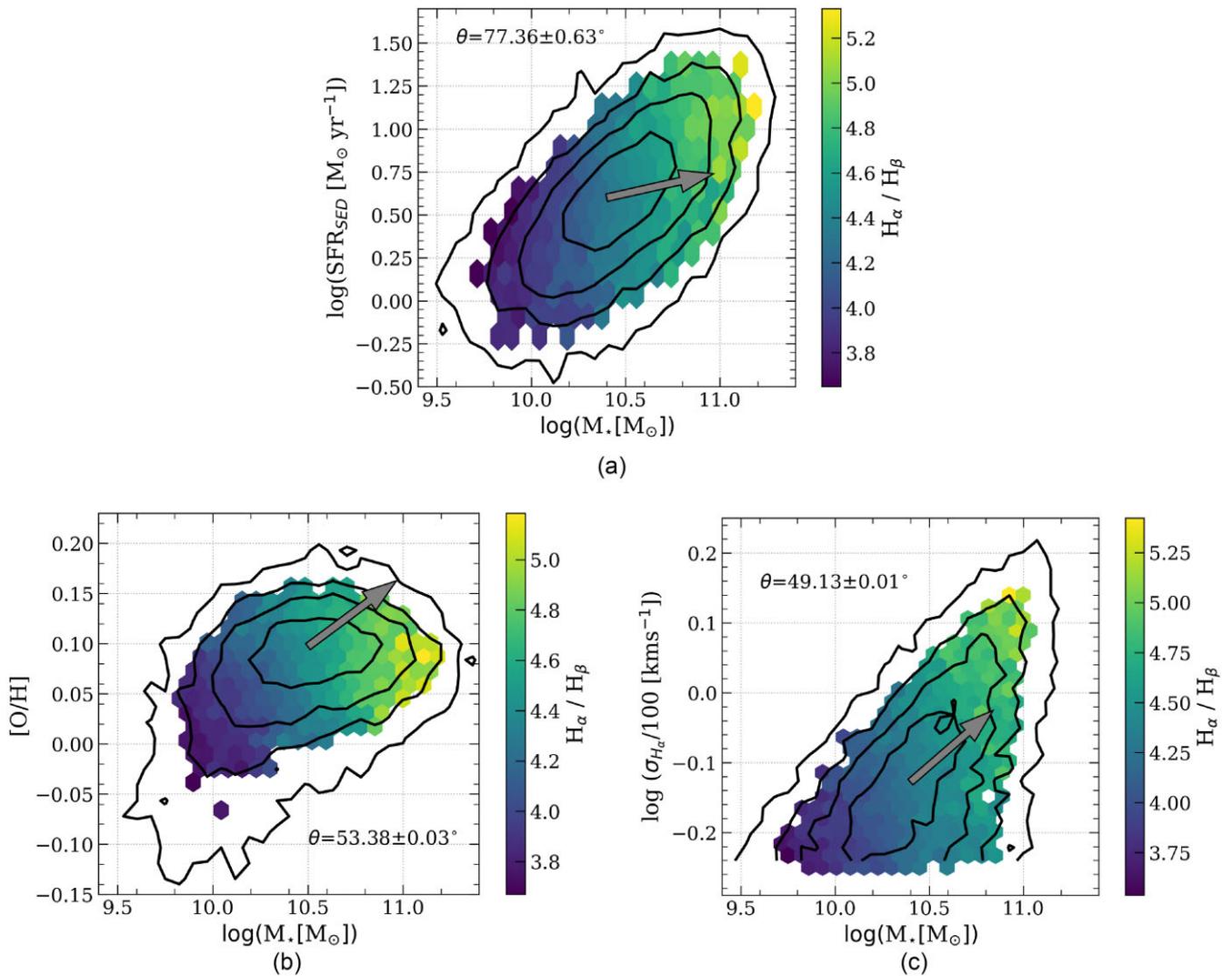

**Figure 2.** SFR (calculated using SED fitting), metallicity and velocity dispersion (normalized by $100\,\mathrm{km\,s^{-1}}$) as a function of stellar mass, colour-coded by the Balmer decrement (i.e. 2D histograms in which the Balmer decrement is the dependent variable), for local galaxies (SDSS). The grey arrows denote the direction in which the Balmer decrement has the largest gradient, determined using the PCC coefficients, with its angle defined clockwise from the positive $y$-axis. The colour gradients and arrows clearly indicate a strong dependence on stellar mass and, at a given stellar mass, also a strong dependence on both metallicity and velocity dispersion, but little or no dependence on SFR. The black contours indicate the density of the galaxies in each diagram, with the outermost contour containing 95 per cent of the galaxies. The contours in (d) do not join due to the sharp cut in $\sigma_{H_\alpha}$.

We additionally see that the PCC between the Balmer decrement and all parameters except the SFR calculated from SED fitting are positive for both samples, implying a positive correlation between the Balmer decrement and those parameters. The PCC between the Balmer decrement and the SFR calculated from SED fitting, however, is negative for the sample with no selection on its inclination, and almost zero for the sample with inclination $i < 45°$, implying any correlation is driven by the inclination, or some other parameter which cross correlates them, and this effect is reduced when the inclination is controlled for.

### 6.1.4 Establishing the analytical dependence of the Balmer decrement on galaxy properties

Combining the results from the RF and PCC analysis, we see that the stellar mass is by far the most important parameter in determining the Balmer decrement. Both the metallicity and the velocity dispersion have significant importance; however, the two analysis methods do not agree on the order of their importance, with RF analysis ranking the velocity dispersion slightly higher than the metallicity, and the PCC analysis ranking the metallicity and the velocity dispersion at similar levels. In this section, we investigate the analytical dependence of the Balmer decrement on these important galaxy properties.

To quantitatively investigate how the Balmer decrement depends on these galactic parameters, we created track plots of the galaxies, with the stellar mass on the $x$-axis, Balmer decrement on the $y$-axis, and the tracks binned in $SFR_{SED}$, metallicity, and nebular velocity dispersion. The tracks represent the stellar mass versus Balmer decrement relation in constant bins of the third variable. We chose the stellar mass to be on the x-axis since it showed the strongest correlation with the Balmer decrement from analysis in the previous sections.

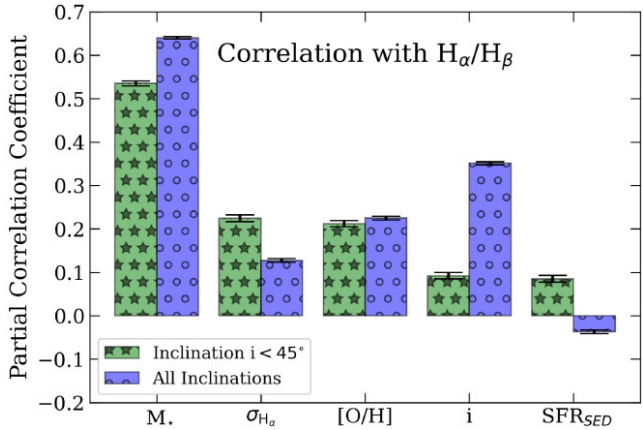

**Figure 3.** PCC of the Balmer decrement with the different galaxy parameters for local galaxies (SDSS). Green bars with stars show the PCC values for galaxies with inclination $i < 45°$, and blue bars with circles show the PCC values for galaxies with all inclinations. The stellar mass is most strongly and intrinsically correlated with the Balmer decrement, in agreement with the RF results, followed by the inclination for the sample of galaxies with no selection on their inclination. Selecting for the inclination $i < 45°$ greatly reduces the PCC value of the Balmer decrement with inclination, and in this case the second strongest Balmer decrement correlation is with metallicity and velocity dispersion, while the correlation with $SFR_{SED}$ (SFR calculated via SED fitting), becomes insignificant.

These track plots are shown in Fig. 4, which illustrates that, while there is always a strong Balmer decrement dependence on the stellar mass, at a fixed stellar mass, the Balmer decrement is dependent most on metallicity and least on SFR. The strength of the Balmer decrement versus metallicity, versus SFR, and versus velocity dispersion dependencies are stellar mass-dependent themselves. The dependence on velocity dispersion is strongest at higher mass, and this trend is inverted for metallicity, with the dependence largest at low masses. There is negligible dependence of the Balmer decrement on the SFR at all masses investigated here, and so is not considered further in the analysis below.

The interdependencies of these galactic parameters are clearly shown here. Following the methodology proposed in Mannucci et al. (2010), we attempted to reduce the dimensionality of the problem by rotating the stellar mass, metallicity and velocity dispersion parameters space such that this projection minimizes the dispersion in the Balmer decrement. This projection reduces the dependence of the Balmer decrement to only one parameter, which we defined as the reduced mass $\mu$,

$$\mu = \log M_\star + \alpha[O/H] + \delta \log \sigma_{100}, \qquad (7)$$

where $M_\star$ is in units of solar mass, and $\sigma_{100}$, is the velocity dispersion (measured from $H_\alpha$) normalized by 100 km s$^{-1}$. This definition and normalization is maintained throughout the rest of this paper. Additionally, $\alpha$ and $\delta$ are parameters that should be determined to minimize the dispersion in the Balmer decrement. The minimization method used is discussed further in Appendix E, where the determined values of $\alpha$ and $\delta$ which minimize the dispersion in the Balmer decrement are 3.67 and 2.96, respectively.

Using these minimization parameters, we recreated Fig. 4 but replacing the stellar mass with the reduced mass $\mu$ at minimum dispersion, as shown in Fig. 5. The tracks in both Fig. 5(a) and (b) are much less spread in Balmer decrement, at a given value of the reduced mass, compared to the tracks in Fig. 4. This shows the dependence of the Balmer decrement on the metallicity and velocity dispersion is being greatly reduced, indicating that our minimization analysis reduced the dimensionality of our problem so that the Balmer decrement depends on one parameter, $\mu$.

To further illustrate the effect of using the reduced mass over the stellar mass, we ran the RF regression on the galaxies with inclination $i < 45°$, whilst including the reduced mass $\mu$ in the analysis as an extra parameter to test whether $\mu$ is now the most important parameter compared to the other global galactic parameters. We also included the un-minimized parameter $\mu_0 = \log M_\star + [O/H] + \log \sigma_{100}$ to test whether the importance of $\mu$ is simply due to the RF picking up on the linear combination of the other parameters, or whether the minimization has had an effect. The importances of each of the parameters are shown in Fig. 6, showing that the importance of $\mu$ in determining the Balmer decrement is dominant, with all other parameters including $\mu_0$ having next to no importance relatively. Hence, the reduced mass encapsulates the majority of the importance of all the other galactic parameters considered in this work.

To show how well this minimization worked on the galaxies themselves, we plot the galaxy contours with the Balmer decrement on the y-axis against (a) the stellar mass and (b) the reduced mass, shown in Fig. 7. In both plots, the mean and error on the mean of the Balmer decrement are shown in blue, and the median and the 84–16th percentile range are shown in green, each calculated in bins 0.15 dex wide of the x-axis parameter. It can be seen that the dispersion or 84–16th percentile range of the Balmer decrement of the galaxies is reduced when moving from having the stellar mass on the x-axis in (a) to the reduced mass on the x-axis in (b). The unweighted average 84–16th percentile range across all the bins in the x-parameter was calculated for each plot, and shown as the red error bar on the plots, with (a) having 0.906 and (b) having 0.849. This shows the effect of the minimization, reducing the percentile range by 6.3 per cent. This result indicates that part of the dispersion in the Balmer decrement versus stellar mass diagram is not intrinsic, but a consequence of the secondary dependences on metallicity and velocity dispersion. Once these dependences are taken into account by introducing $\mu$, then the scatter is reduced. The residual scatter is likely due to diverse evolutionary processes within the galaxies, although it may also partly be due to observational errors. The contribution of the observational errors to the overall scatter we see was estimated by taking the median error on the measurement of the Balmer decrement across the sample used in this work, with the observational error calculated to be 0.32. The reduction in the percentile range is small, although this is not surpizing since the stellar mass accounted for the majority of the variation in the Balmer decrement, so accounting for the less important (but still important) parameters would have a small, non-negligible effect.

In order to provide a functional form of the Balmer decrement versus stellar mass and versus. reduced mass dependence, we fit a third-order polynomial to the mean of the Balmer decrement in each of those parameter spaces, providing the following fits:

$$\begin{aligned}
H_\alpha/H_\beta = &(-0.193 \pm 0.067) \log(M_\star[M_\odot])^3 \\
&+ (6.097 \pm 2.059) \log(M_\star[M_\odot])^2 \\
&+ (-63.163 \pm 21.236) \log(M_\star[M_\odot]) \\
&+ (218.705 \pm 72.946) \qquad (8)
\end{aligned}$$

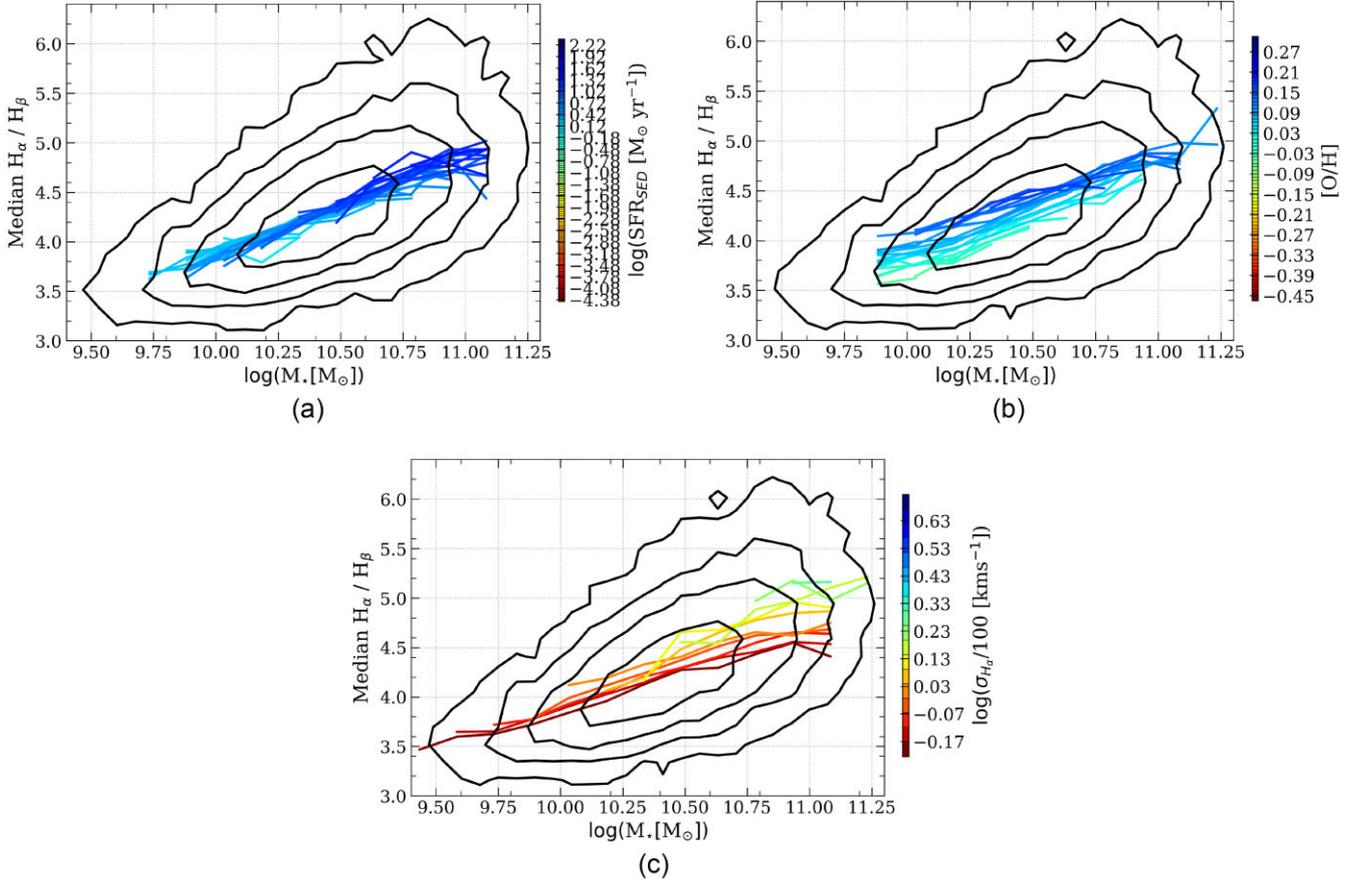

**Figure 4.** Balmer decrement as a function of stellar mass in bins of (a) SFR (calculated via SED fitting), (b) metallicity, and (c) velocity dispersion. These tracks confirm that, at a fixed stellar mass, the Balmer decrement depends on metallicity and velocity dispersion, but has negligible dependence on SFR. The contours in black display the density of galaxies in each diagram, with the outermost contour containing 95 per cent of the galaxies.

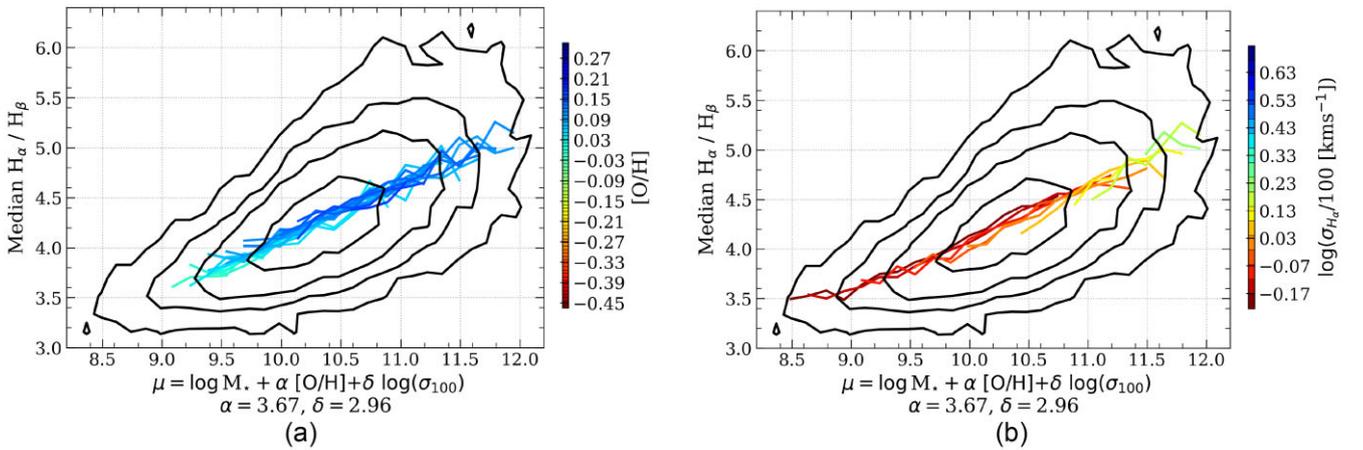

**Figure 5.** Balmer decrement versus reduced mass $\mu = \log M_* + \alpha[O/H] + \delta \log \sigma_{100}$, where $\alpha = 3.67$ and $\delta = 2.96$, in bins of metallicity (left panel) and velocity dispersion (right panel). The fact that, at a fixed $\mu$, there is little/no dependence on the Balmer decrement on either metallicity or velocity dispersion indicates that the reduced mass has captured well these secondary dependences. In particular, compared to the track plots in Fig. 4, the dependence of the Balmer decrement on the colour-coded parameters is considerably reduced. The contours in black display the density of galaxies in these diagrams, with the outermost contour containing 95 per cent of the galaxies.

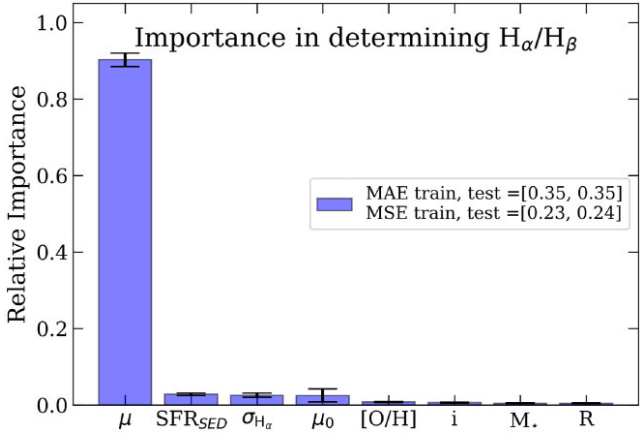

**Figure 6.** Importance of the various galactic parameters and the reduced mass $\mu$ in determining the Balmer decrement, for local galaxies (SDSS) with inclination $i < 45°$, as inferred with the RF analysis. The reduced mass $\mu$ is now by far the most important parameter, reducing the importance of the stellar mass, metallicity and velocity dispersion to be negligible, meaning all importance of these three galactic parameters are well incorporated in the reduced mass $\mu$ for what concerns their role in determining the Balmer decrement. The un-minimized parameter $\mu_0 = \log M_\star + [O/H] + \log \sigma_{100}$ is also included to show the importance of the minimized $\mu$ is not simply due to the RF regressor picking up the linear combination of the other parameters. The difference between the average MAE and MSE of the train and test samples being so small implies no overfitting. Parameter R is the random variable and SFR$_{SED}$ is the SFR calculated using SED fitting.

and

$$H_\alpha/H_\beta = (-0.027 \pm 0.005)\mu^3 + (0.848 \pm 0.160)\mu^2$$
$$+(-8.385 \pm 1.608)\mu + (29.856 \pm 5.368) \quad (9)$$

and the resulting fits are shown in Fig. 7 as the orange lines.

In order to further demonstrate that the reduced mass $\mu$ encapsulates all the variation of the Balmer decrement due to indirect correlations with the other galactic parameters (metalliicty and velocity dispersion), we recreate Fig. 2 by plotting both the metallicity and velocity dispersion now as functions of the reduced mass $\mu$, colour-coded by the Balmer decrement, as is shown in Fig. 8. When replacing the stellar mass with the reduced mass $\mu$, the dependence of the Balmer decrement on both the metallicity and the velocity dispersion (whilst controlling for the reduced mass) greatly reduces, as it can be seen by eye or by considering the arrow angles, which each rotate to within about 20° or horizontal (90°). Hence, using this projected parameter space encapsulates the dependence of the Balmer decrement on a single parameter.

### 6.1.5 Summary

We have identified the most important parameters in determining the Balmer decrement through RF and PCC analysis, finding the stellar mass to be the most important, followed by metallicity and nebular velocity dispersion (once the dependence on the inclination, $i$, is removed by selecting galaxies with $i < 45°$).

The strong dependence on stellar mass is in line with the expectation from equation (4), where this dependence primarily comes from the MGMS. The dependence on metallicity is also in line with equation (4), where this dependence comes from the relationship between the dust-to-gas ratio and the metallicity, since DGR = DZR*$Z_g$.

The additional dependence on the nebular velocity dispersion was not expected. This may be due to how the nebular velocity dispersion traces the gravitational potential in the galaxy, which is the capability of the galaxy to retain dust against the strong radiation pressure on the dust and to retain metals against metal loss via winds and gas outflows (Chisholm et al. 2015).

Additionally, our analysis has determined the SFR derived from $H_\alpha$ is solely important due to the $H_\alpha$ flux used in calculating the SFR being dust corrected itself, and the SFR calculated using SED fitting is a more valid tracer of the SFR in this work. This tracer of the SFR is shown to be not important in determining the Balmer decrement when compared to the stellar mass, metallicity and velocity dispersion.

By combining these important parameters into the reduced mass $\mu$, we have been able to collapse the majority of the dependence of the Balmer decrement to this parameter. This will allow for much easier comparison with other samples of galaxies in the next section.

Quantitatively, the dependence on stellar mass and metallicity is in the right direction, however, does not match exactly with the expectations from the simple predictions in equation (4). To better predict these observations, more advanced modelling would be required, including a comparison with numerical simulations and potentially considering a geometrical factor dependent on mass, which is currently assumed constant. Zuckerman et al. (2021) argue that the dust attenuation is related to the thickness of the galaxy, and since a galaxy with higher stellar mass will have a greater thickness, it is likely that the disc thickness contributes some of the correlation between stellar mass and dust attenuation we observe.

### 6.2 Comparison of samples at high redshifts

Although the focus of this paper is primarily to investigate the scaling relations between dust attenuation (traced by the Balmer decrement) and galaxy properties in the local universe, it is interesting to explore also whether such relations hold at high redshift. Samples at high redshift have much lower statistics and higher uncertainties; hence, the level of analysis performed in this paper on the local sample is certainly not possible on high redshift samples, at least not yet. However, we can explore whether their properties are consistent with the local findings.

Specifically, we investigate whether the scaling relations given by equations (8) and (9) that we have found in the local Universe between dust attenuation and other global galactic properties hold at high redshift. We do this by comparing the observed values of the Balmer decrement from the galaxies in the KLEVER and MOSDEF surveys ($z \sim 1-3$) with the galaxies in the SDSS survey in stellar mass space and in reduced mass space, which should encapsulate all of the parameters important in determining the Balmer decrement. As mentioned previously, new samples at even higher redshift from NIRSpec-JWST surveys do not yet have the information required to perform these tests.

Plots comparing the local and higher redshift galaxies are shown in Fig. 9, where the Balmer decrement is plotted against stellar mass and also against reduced mass for the galaxies in the KLEVER survey (a) and (b), and the galaxies in the MOSDEF survey (c) and (d). Here the mean and error on the mean were calculated in order to focus on the primary dependences. For both the galaxies in KLEVER and MOSDEF, this shows the Balmer decrements to overlap with those of the local galaxies in both the stellar mass and reduced mass space. These findings are consistent with no redshift evolution of these relations. For the simple dependence on mass, this finding agrees with the results from Shapley et al. (2022) for the galaxies in MOSDEF, and with the results from Shapley et al. (2023) at even higher redshifts.

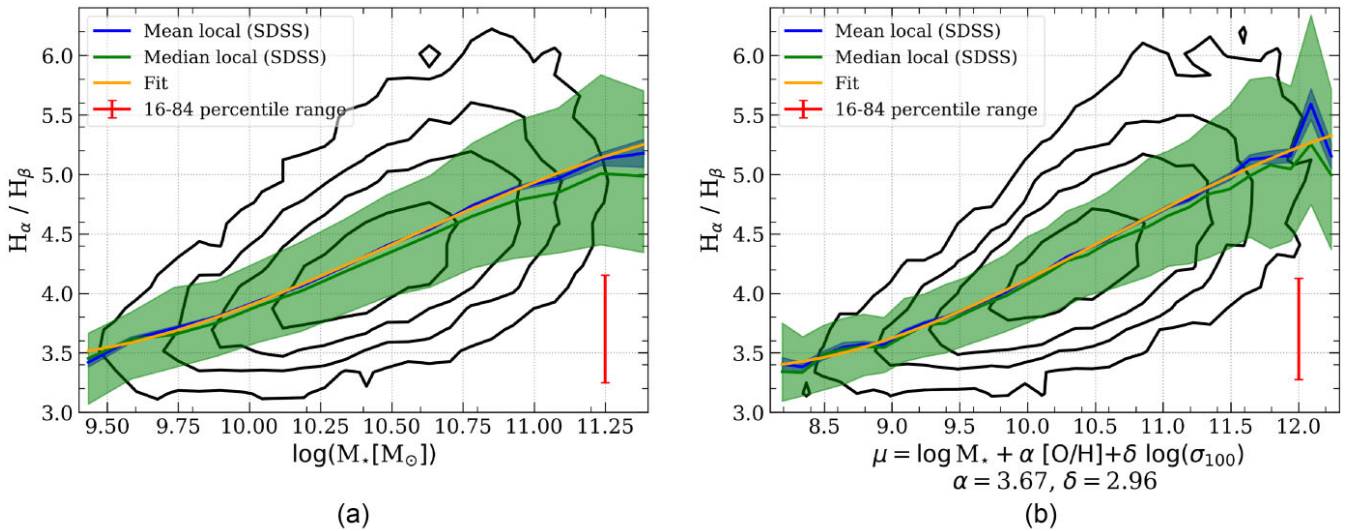

**Figure 7.** Contours showing the distribution of the Balmer decrement of local galaxies (SDSS) as a function of (a) stellar mass and (b) reduced mass. The red error bars represent the average 16–84th percentile range of the Balmer decrement. The blue line represents the mean Balmer decrement in bins of 0.15 dex in the *x*-axis quantity; the shaded blue region represents the error on the mean in each bin. The green lines represent the median Balmer decrement, and the green-shaded region the 84–16th percentile range in each bin. The orange line represents the third-order polynomial fit to the mean. The outermost contour contains 95 per cent of the galaxies.

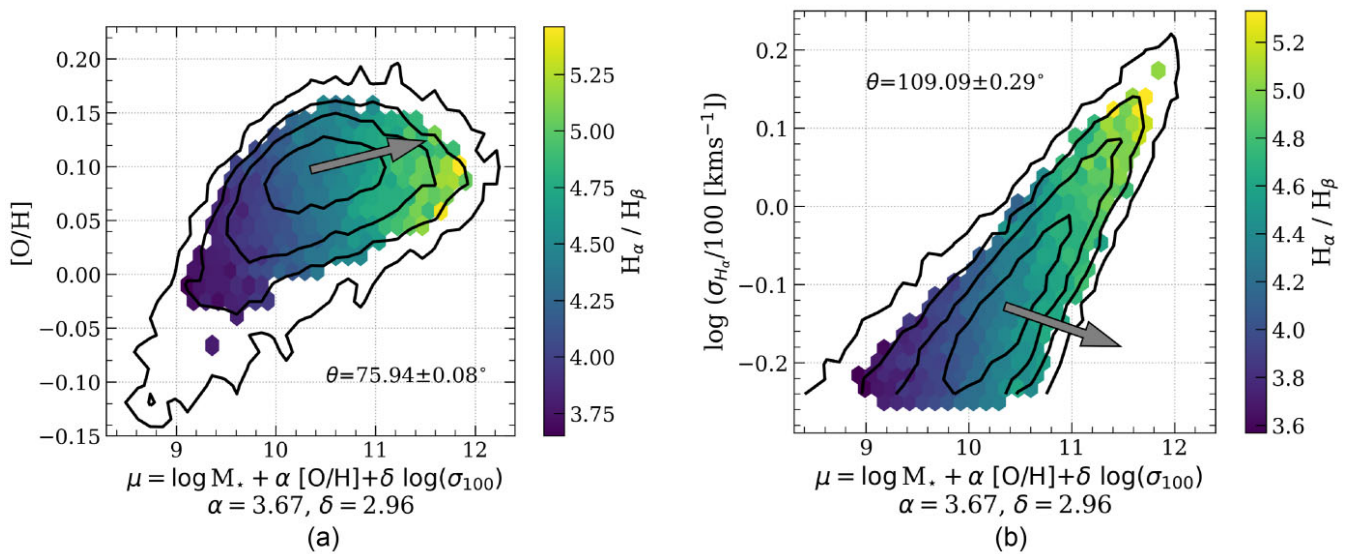

**Figure 8.** Metallicity and velocity dispersion as a function of reduced stellar mass $\mu$, colour-coded by Balmer decrement (i.e. 2D histograms in which the Balmer decrement is the dependent variable), for local galaxies (SDSS). The grey arrows denote the direction in which the Balmer decrement has the largest gradient, determined using the PCC coefficients. The colour gradients and arrows clearly indicate an even stronger dependence of the Balmer decrement on reduced stellar mass $\mu$, relative to the dependence on the stellar mass seen in Fig. 2, while the dependence of the Balmer decrement on metallicity and velocity dispersion is now greatly reduced with respect to Fig. 2 (the residual dependence is due to the fact that these diagrams explore only the dependence of two quantities at each time, hence they pick the residual dependence on all other quantities). The black contours indicate the density of the galaxies in each diagram, with the outermost contour containing 95 per cent of the galaxies. The contours in (b) do not join due to the sharp cut in $\sigma_{H_\alpha}$.

The Balmer decrement versus stellar mass relationship for the galaxies in the MOSDEF survey seems to slightly flatten compared to galaxies in the SDSS survey, which can be explained by the fact the MOSDEF survey is complete for stellar masses less than $10^{10.5} M_\odot$, however, incomplete for dusty red SFGs with stellar mass above $10^{10.5} M_\odot$. Hence, the dustiest massive galaxies might be missed, causing the Balmer decrement versus stellar mass relationship to flatten at high stellar masses.

This lack of evolution in the stellar mass space could be due to a truly redshift invariant Balmer decrement versus stellar mass relationship. As suggested by Shapley et al. (2022), a non-evolving relation can arise due to offsetting effects from the simultaneous evolution of gas mass surface density, DGR, metallicity, dust geometry, and/or dust mass absorption coefficients. Our finding of no evolution even in the Balmer decrement relation with *reduced* mass $\mu$, makes the explanation of a combination of different evolutionary effects cancelling each other unlikely. Our findings are more supportive of a scenario in which the dust production mechanism and associated distribution in galaxies do not change with cosmic time up to $z \sim 2$–3, i.e. the multidimensional relationship between dust attenuation and the galactic quantities does not change with cosmic epoch, galaxies simply populate different regions of this multidimensional surface at different cosmicepochs.

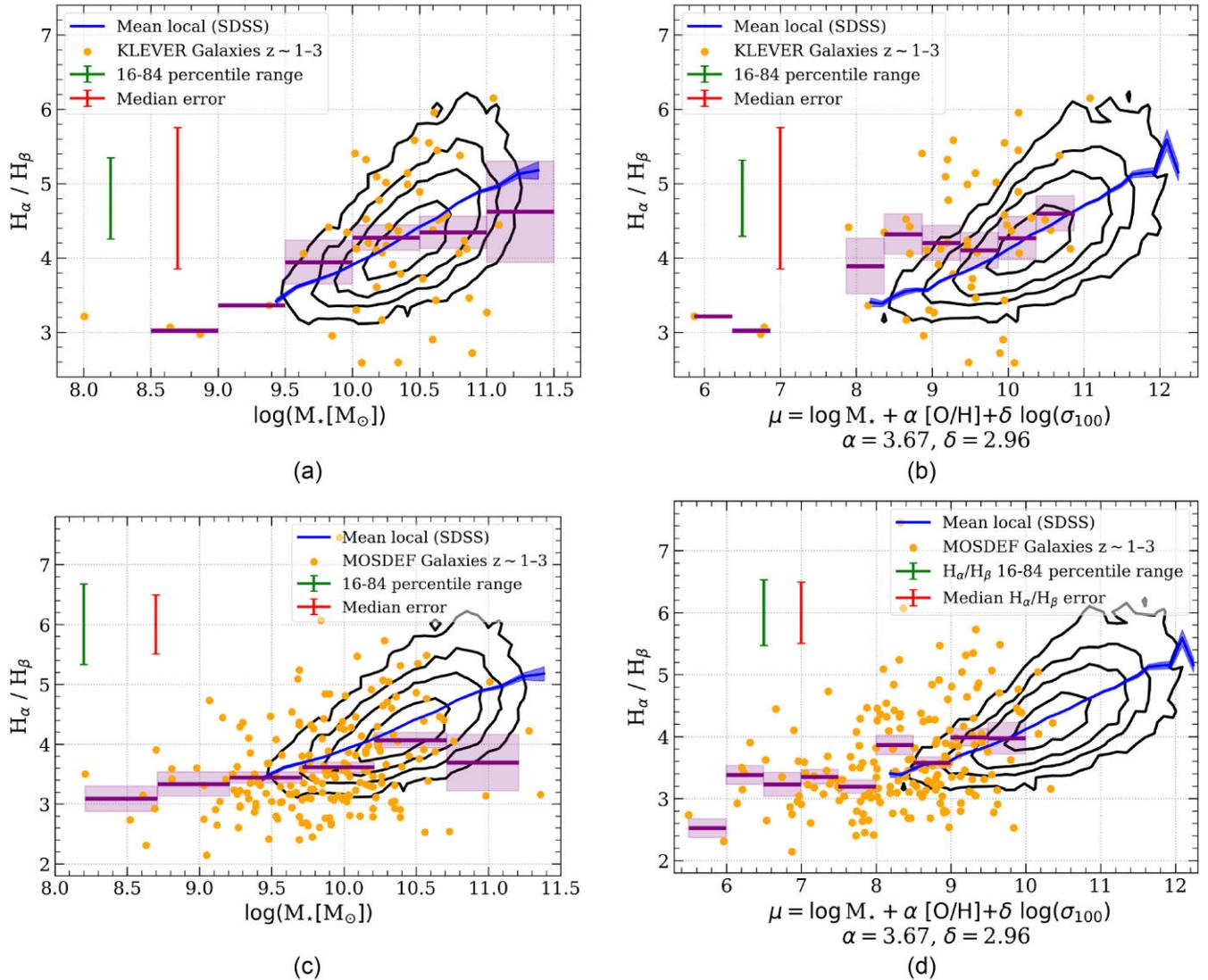

**Figure 9.** Balmer decrement as function of stellar mass (left panel) and reduced mass (right panel) comparing local galaxies (SDSS) with galaxies at high redshift ($z \sim 1$–3) from the KLEVER survey (top panel) and the MOSDEF survey (bottom panel). The distribution of local galaxies in SDSS is shown by the black contours, where the outermost contour contains 95 per cent of the galaxies, and the blue line is the mean Balmer decrement in each 0.15 dex wide bin in stellar mass or reduced mass with at least 25 galaxies present. Shaded blue regions represent the error on the mean in each bin. High-$z$ galaxies in the KLEVER and MOSDEF surveys are shown in orange. The purple segments show the means and the purple-shaded regions the errors on the mean for each 0.5 dex wide bin in stellar mass and reduced mass. The green error bars represent the average 16–84th percentile ranges and the red error bars represent the median error in the Balmer decrement measurement for the high-$z$ galaxies. Panels (a) and (b) show that there is no significant evolution between the Balmer decrement of the galaxies from KLEVER and from the local galaxies both in stellar mass and reduced mass space. Similarly, panels (c) and (d) show that there is no significant evolution between the Balmer decrement of the galaxies from MOSDEF and the local SDSS galaxies in both stellar mass and reduced mass space. These results indicate that there is no evolution of the relation between the Balmer decrement versus stellar mass and of the Balmer decrement versus reduced mass up to a redshift of $z \sim 1$–3. The slight flattening of the Balmer decrement stellar mass relation at high masses ($M_\star > 10^{10.5}\,\mathrm{M}_\odot$) might be due to the MOSDEF survey possibly missing a portion of the massive dusty red SFGs.

The inclination of the $z \sim 1$–3 galaxies was not controlled for in this work. Lorenz et al. (2023) show that for their sample of galaxies taken from the MOSDEF survey ($1.37 \leq z \leq 2.61$), there is no dependence of the dust attenuation on inclination. They propose a dust model where attenuation occurs in three components: star-forming regions, large dusty star-forming clumps, and a small contribution from the ISM. Since the majority of the attenuation occurs in the roughly spherical star-forming regions and large dusty star-forming clumps, there would be no dependence of the attenuation on the inclination of the galaxy. This three-component model is similar to the widely used two-component model for local galaxies ($z \sim 0$), consisting of star-forming regions with optically thick dust primarily around young stars and then the diffuse ISM component (Charlot & Fall 2000). The addition of the large dusty star-forming clumps in the three-component model is supported by observations of ($z \sim 2$) galaxies from, for example, Schreiber et al. (2011) and Wuyts et al. (2012). This model also supports the dust attenuation versus stellar mass relationship, since these large dusty star-forming clumps have been observed to be both larger Swinbank et al. (2012) and more common (Tadaki et al. 2013) as stellar mass increases, and a larger dusty star-forming clump will lead to increased dust attenuation due to extended path-lengths for the light.

The analysis in this work suggests that the dust attenuation of our local sample of galaxies (SDSS) is dependent on the inclination of the galaxies, and this result can be supported by the two-component dust model. Our statistical analysis of the galaxy parameters is picking up on the dependence between dust attenuation and the inclination of the galaxy, likely caused by the attenuation originating in the diffuse ISM.

We conclude, however, by warning that the comparison with high-redshift galaxies is still plagued by the large dispersion and poor statistics, which makes even the errors on the means still relatively large (as highlighted in Fig. 9), and a more thorough exploration requires much larger samples, which may become available with the next-generation near-IR MOS spectrographs (Maiolino et al. 2020).

# 7 CONCLUSION

In this work, we have investigated which galactic parameters are most important in determining the dust attenuation in galaxies, as traced by the Balmer decrement, and exploring how this varied at different cosmic epochs by comparing local galaxies (SDSS) with samples at $z \sim 1-3$ (KLEVER and MOSDEF).

We summarize our results as follows:

(i) PCC and RF analysis on local (SDSS) galaxies show that the stellar mass is the most important parameter in determining the dust attenuation traced by the Balmer decrement. Metallicity and nebular velocity dispersion are also important but less so than the stellar mass.

(ii) Galaxy inclination has obviously an important effect on the observed attenuation. However, its effect on these results was controlled for by selecting galaxies with inclination $i < 45°$; with this selection, the Balmer decrement had negligible dependence on the inclination via PCC and RF analysis.

(iii) The dependence of the Balmer decrement on SFR traced by $H_\alpha$ is driven by the fact that $H_\alpha$ is also included in the Balmer decrement, and by the fact that the Balmer decrement being used to dust correct the $H_\alpha$ flux; hence, the correlation between Balmer decrement and SFR inferred from $H_\alpha$ is spurious. No dependence of the Balmer decrement on SFR is found if the latter is inferred using SED fitting.

(iv) The dispersion of the Balmer decrement in the rotated parameter space defined by the reduced mass, $\mu = \log M_\star + 3.67[O/H] + 2.96 \log \sigma_{100}$, is reduced compared to the dispersion in stellar mass space. This indicates that the variation in the Balmer decrement due to the metallicity and velocity dispersion are captured by this reduced mass.

(v) The dependence of the Balmer decrement on the stellar mass is expected from the MGMS relation ($M_{H2}$ versus $M_\star$). The dependence on metallicity is also expected from the dust-to-gas ratio. The dependence on velocity dispersion was not expected and may trace the capability of more massive systems (traced by the higher velocity dispersion) to better retain dusty clouds against radiation-driven pressure outflows.

(vi) We observe no significant evolution of the relationship between the Balmer decrement and stellar mass relationship up to $z \sim 1-3$. Hence, the dust attenuation versus stellar mass relationship does not evolve up to this redshift. We additionally see no significant evolution of the relationship between the Balmer decrement and the reduced mass, $\mu$, indicating that the scaling relations found locally capture the dust attenuation properties also of distant galaxies.

This work can be greatly expanded at high redshift with the next generation, large multiplexing near-IR spectrographs (e.g. MOONS: Cirasuolo et al. 2020; Maiolino et al. 2020), which will provide spectra for several hundred thousands galaxies, i.e. with statistics similar to the SDSS, around cosmic noon ($z \sim 1-3$).

This work can also be extended to higher redshifts using data from JWST's NIRSpec surveys and NIRCam slitless mode. This exploration has already started for what concerns the dependence of the Balmer decrement as a function of stellar mass (Shapley et al. 2023), but can be expanded further to also investigate the relation with the reduced mass. JWST spectroscopic surveys are expected to detect thousands of galaxies out to $z \sim 7$, for which both $H_\alpha$ and $H_\beta$ will be available. Hence, it will be possible to investigate the Balmer decrement across a very large range of redshifts and track the evolution of the dust attenuation versus reduced mass relation.

## ACKNOWLEDGEMENTS

GM and RM acknowledge support by the Science and Technology Facilities Council (STFC), by the ERC through Advanced Grant 695 671 'QUENCH' and by the UKRI Frontier Research grant RISEandFALL. RM also acknowledges funding from a research professorship from the Royal Society.

## DATA AVAILABILITY

The MPA-JHU catalogue is publicly available at https://skyserver.sdss.org/dr12/en/help/docs/tabledesc.aspx?name = galSpecInfo. The SDSS morphological catalogue is publicly available at https://cdsarc.cds.unistra.fr/viz-bin/cat/J/ApJS/196/11. The SDSS SFRs calculated through SED fitting are publicly available at https://cdsarc.cds.unistra.fr/viz-bin/cat/J/ApJS/219/8. The MOSDEF catalogue is publicly available at https://mosdef.astro.berkeley.edu/for-scientists/data-releases/. The KLEVER data are publicly available and attached to its survey paper (Curti et al. 2020b).

# APPENDIX A: LOCAL SAMPLE ANALYSIS WITH SFR$_{H_\alpha}$

Repeating the analysis from Section 6.1.2 using SFR$_{H_\alpha}$, we produce the 2D histogram visualization shown in Fig. A1, demonstrating a stronger correlation between $SFR_{H_\alpha}$ and the Balmer decrement than between the SFR derived from SED fitting and the Balmer decrement shown in Fig. 2.

Additionally, the results in Sections 6.1.1 and 6.1.3 are repeated whilst including the SFR derived from the $H_\alpha$ emission-line flux. These are shown in Fig. A2, with the RF importances in determining the Balmer decrement for each parameter shown on the left (a), and the PCC with the Balmer decrement for each parameter shown on the right (b). In each plot, the blue bars with circles indicate the results using the sample of local galaxies with no selection on their inclination, and the green bars with stars represent the results using the sample of local galaxies with inclination $i < 45°$. Again, both these results show that the relationship between the Balmer decrement and $SFR_{H_\alpha}$ is now much stronger than the relationship with $SFR_{SED}$.

These results indicate that this strong relationship between $SFR_{H_\alpha}$ and Balmer decrement is a spurious artefact of the fact that $H_\alpha$ is a quantity also included in both quantities; moreover, $SFR_{H_\alpha}$ is corrected for dust attenuation using the Balmer decrement itself, hence introducing another spurious correlation.

We also show how the SFR derived from SED fitting relates to the SFR derived from the $H_\alpha$ emission-line flux in Fig. A3, where the straight line is fit using Orthogonal Distance Regression (ODR), estimating the slope to be $0.996 \pm 0.038$, with the scatter being $0.132$ dex, measured as the square root of the residuals in the fit.

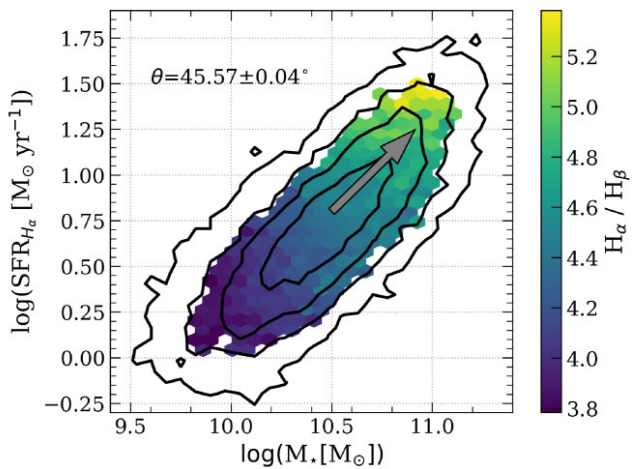

**Figure A1.** SFR (estimated from the $H_\alpha$ flux as a function of stellar mass, colour-coded by Balmer decrement for local galaxies (SDSS). The grey arrow denotes the direction in which the Balmer decrement has the largest gradient, determined using the PCC coefficients, with its angle defined clockwise from the positive *y*-axis. The colour gradients and arrows clearly indicate a strong dependence of the Balmer decrement on both the stellar mass and, at a given stellar mass, also a strong dependence on $SFR_{H_\alpha}$. Comparing to Fig. 2, the correlation between the Balmer decrement and $SFR_{H_\alpha}$ is much larger than with the SFR derived from SED fitting. The black contours indicate the density of the galaxies in each diagram, with the outermost contour containing 95 per cent of the galaxies.

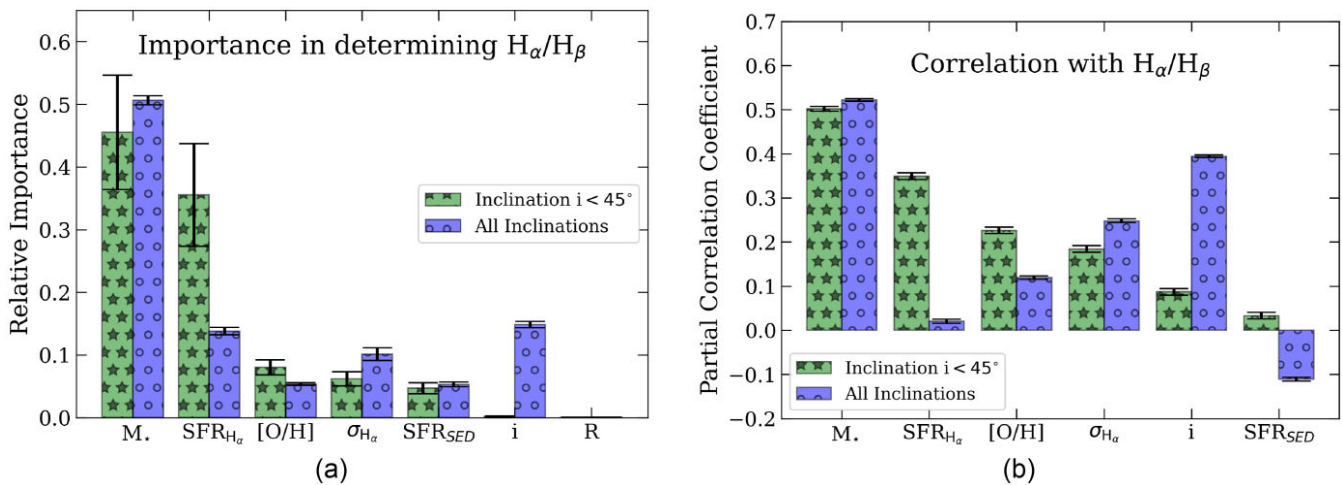

**Figure A2.** Plots showing (a) the importances of the different galactic parameters in determining the Balmer decrement and (b) the PCCs between the Balmer decrement and the different galactic parameters, along with their errors for local galaxies (SDSS), including the SFR derived from $H_\alpha$. Green bars with stars show results for galaxies with inclination $i < 45°$, and blue bars with circles show results for galaxies with all inclinations. Discrepancy between results for $SFR_{H_\alpha}$ and the SFR derived from SEDs highlight how the stronger relationship between the Balmer decrement and $SFR_{H_\alpha}$ is due to the cross-correlation from the Balmer decrement being used to dust correct the $H_\alpha$ line flux. Parameter *R* is the random variable and $SFR_{SED}$ is the SFR derived SED fitting.

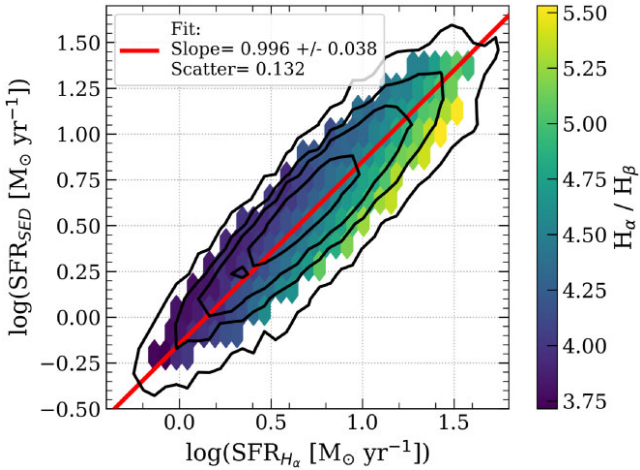

**Figure A3.** SFR estimated from SED fitting as a function of the SFR estimated from the $H_\alpha$ flux, colour-coded by Balmer decrement for local galaxies (SDSS). The red line represents the fit using ODR, showing a near one-to-one slope, and with scatter (taken as the square root of the residuals in the fit) of 0.132 dex. The black contours indicate the density of the galaxies in each diagram, with the outermost contour containing 95 per cent of the galaxies.

## APPENDIX B: D4000–SFR CALIBRATION AND LOCAL SAMPLE ANALYSIS WITH SFR$_{D4000}$

The 4000-Å break is defined as the ratio of flux either side of the break in rest-frame galaxy spectra observed at 4000 Å, and acts as a tracer of the young and old stars in a galaxy spectrum.

In this work, we follow the narrow 4000-Å break (D4000) definition from Balogh et al. (1999), as it is less affected by dust reddening:

$$D4000 \equiv \int_{4000}^{4100} f_\nu \, d\lambda \Big/ \int_{3850}^{3950} f_\nu \, d\lambda \quad , \qquad (B1)$$

where $f_\lambda$ is the spectral flux density of the galaxy in Å.

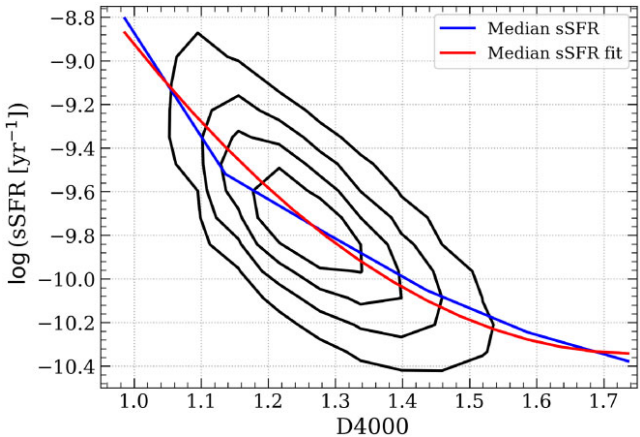

**Figure B1.** Plot of contours of galaxies in the SDSS survey in D4000 and sSFR space, with median sSFR overplotted in blue and the best-fitting 2D polynomial in red. The median sSFR was determined using 0.15 dex wide bins in D4000 which had at least 25 galaxies in. This calibration will be used to determine the SFR of galaxies in the SDSS survey without directly using $H_\alpha$.

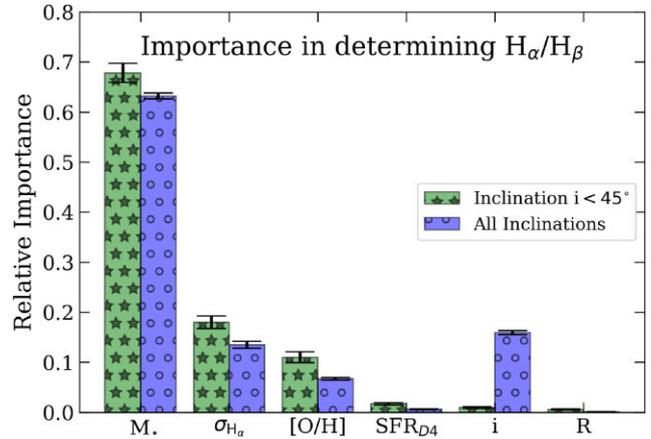

**Figure B2.** Relative importance of the different galactic parameters in determining the Balmer decrement for local galaxies (SDSS) calculated using RF. Green bars with stars show importances for galaxies with inclination $i < 45°$, while blue bars with circles show the importance for galaxies with all inclinations. Stellar mass is the most important parameter in both samples. Selecting for the inclination $i < 45°$ reduces the importance of the inclination to be negligible. SFR$_{D4}$ is the SFR derived from D4000. The parameter $R$ is a control, random variable.

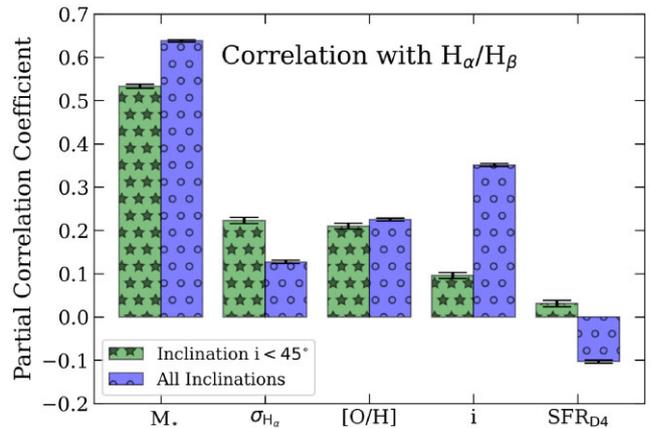

**Figure B3.** PCC of the Balmer decrement with the different galaxy parameters for local galaxies (SDSS). Green bars with stars show the PCC values for galaxies with inclination $i < 45°$, and blue bars with circles show the PCC values for galaxies with all inclinations. The stellar mass is most strongly and intrinsically correlated with the Balmer decrement, in agreement with the RF results, followed by the inclination for the sample of galaxies with no selection on their inclination. Selecting for the inclination $i < 45°$ greatly reduces the PCC value of the Balmer decrement with inclination, and in this case the second strongest Balmer decrement correlation is with metallicity and velocity dispersion, while the correlation with SFR$_{D4}$ (SFR derived from D4000), becomes insignificant.

This empirical correlation is used to determine the SFR for galaxies where measurements on $H_\alpha$ are not available, as is done in Brinchmann et al. (2004) and Bluck et al. (2020a).

In this work, the method proposed by Bluck et al. (2020a) is adopted wherein a calibration is calculated between D4000 and the sSFR, taking the median sSFR in each 0.15 dex wide bin of D4000 with at least 25 galaxies present. This calibration is shown in Fig. B1.

Additionally, the SFR calculated using D4000 was used as the tracer of SFR in the statistical analysis of the local galaxies as is done in Section 5, where instead the SFR from SED fitting was used. Analysis using RF and PCCs to determine the importance of the different parameters in determining the Balmer decrement are shown in Figs B2 and B3, respectively. Both figures show similar results to those shown in Figs 1 and 3 in the main text, where the SFR is calculated instead using SED fitting.

The result of using the SFR from D4000 instead of the SFR from SED fitting implies the same conclusion, that the dependence of the Balmer decrement on the SFR traced by the $H_\alpha$ flux is spuriously high due to the $H_\alpha$ flux being itself dust corrected using the Balmer decrement, as well as $H_\alpha$ being present in the Balmer decrement itself.

## APPENDIX C: RF TUNING

We train an RF regressor to each of the local galaxy samples, first with no selection on their inclination and second with inclination $i < 45°$, to determine which parameters are most important in determining the Balmer decrement. First, we tuned the minimum number of samples on the final leaf hyperparameter. To do this, we varied this hyperparameter and trained an RF regressor with each hyperparameter, then determined the mean absolute error (MAE) and mean-squared error (MSE) between the input and predicted Balmer decrements for both the train and then test sample. The MAE and MSE are metrics of the prediction accuracy of the regressor, i.e. how well the regressor predicts the feature parameter. The optimal hyperparameter is when the difference between the train and test MAE is below 2 per cent, implying no overfitting. Overfitting occurs when the RF is fit too well to the training data that it can not adapt to new test data, leading to a large difference between the prediction accuracy for the train and test samples.

This method was repeated five separate times on each sample, and the averaged MAE at each value of the hyperparameter is shown in Fig. C1(a) for the galaxies with no selection on their inclination, and in (b) for the galaxies with inclination $i < 45°$. To additionally check that no overfitting occured in the RF, we plotted the input against predicted Balmer decrement for the train and then test sample, as is shown in Figs C1(b) and (d) for the two samples, where the orange contours represent the training data set, and the blue the testing data set. Since we can see no significant difference between the train and test sample for both samples, the two regressors are not overfit. Using the determined best hyperparameter value for each regressor, we ran the RF 100 times on their respective galaxy sample and averaged the importances for each parameter, and took the standard deviation of the importances to represent their error.

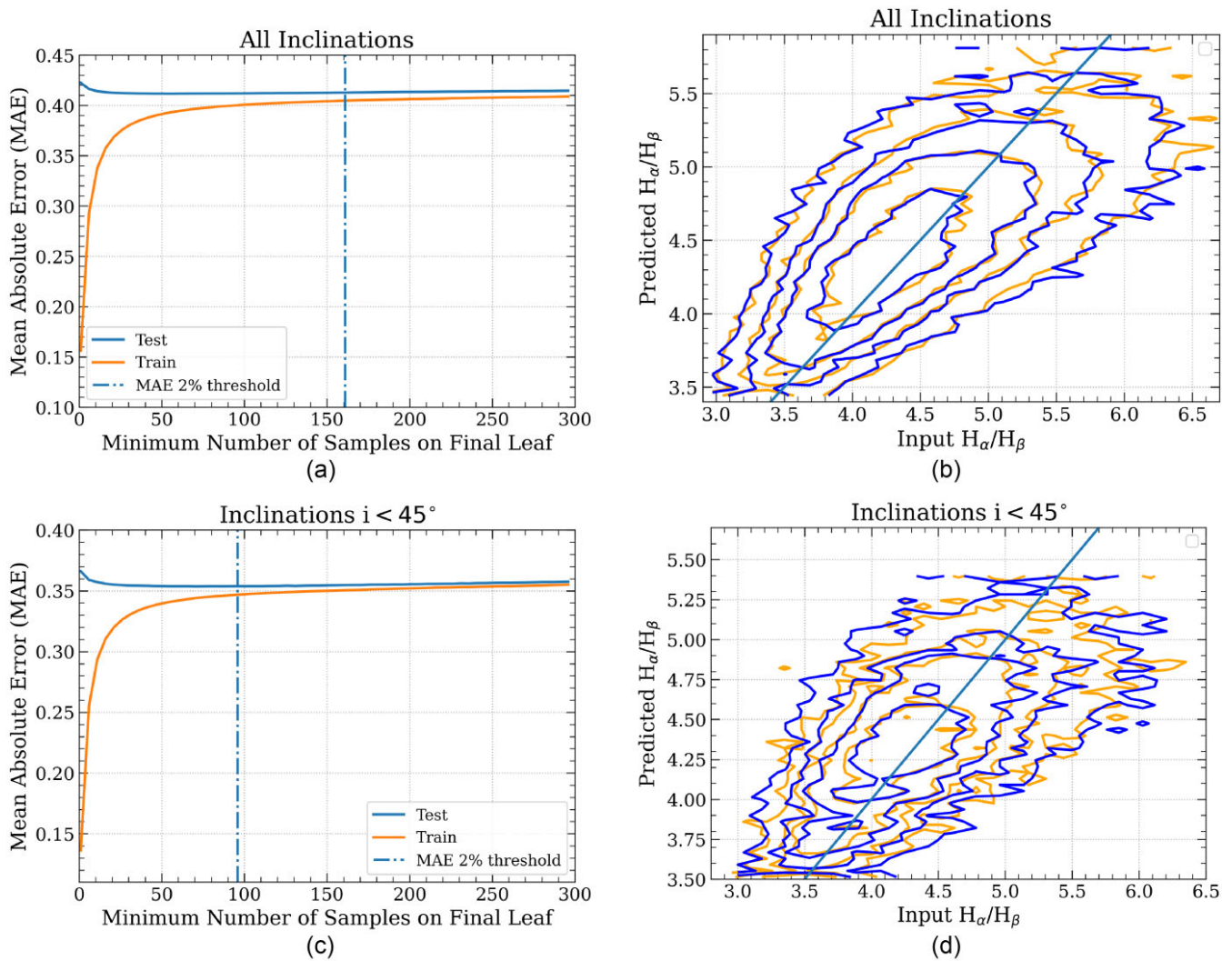

**Figure C1.** Plots showing the tuning performed on the local galaxies with no selection on their inclination, (a) and (b), and on the local galaxies with inclination $i < 45°$, (c) and (d), used to produce the importances in Fig. 1. Plots (a) and (c) show the minimum number of samples on the final leaf plotted against the average MAE over five iterations for the train and the test sample. The optimal hyperparameter value is the minimum value at which the difference between the train and test MAE is less than 2 per cent. Plots (b) and (d) show the contours of the input Balmer decrement plotted against the predicted Balmer decrement for the train (orange) and test (blue) samples, with a 1:1 line in blue, showing the regressor is not overfitting for both samples due to a strong overlap of the two sets of contours. The outermost contour of each sample contains 95 per cent of the galaxies.

## APPENDIX D: STELLAR VELOCITY DISPERSION

The dependence of the Balmer decrement on the stellar velocity dispersion was additionally investigated in this work for the local galaxies.

To use the stellar velocity dispersion ($\sigma_\star$), we first explored how it related to the nebular velocity dispersion traced by $H_\alpha$ emission ($\sigma_{H_\alpha}$) by plotting the nebular velocity dispersion as a function of stellar velocity dispersion, with the Balmer decrement on the $z$-axis. This plot is shown in Fig. D1, using the local galaxies from the SDSS survey cut such that their inclination is less than 45°. The angle of the PCC arrow (44°.31) implies the dependence of the Balmer decrement on the two different velocity dispersions is almost equal. However, the galaxies are offset from the grey line representing the 1:1 relation between the two velocity dispersions, indicating that the two velocity dispersion are not completely interchangeable in this analysis. As is shown for $z > 1$ galaxies (Übler et al. 2022) and for local galaxies (Crespo Gómez et al. 2021), the stellar velocity dispersion is, on average, twice the nebular velocity dispersion.

To further understand if we could use the stellar velocity dispersion in our analysis of local galaxies instead of the nebular velocity dispersion, we ran RF regression on our sample of galaxies with inclination less than 45° as is done in the main text, however, replacing $\sigma_{H_\alpha}$ with $\sigma_\star$. The RF regressor was tuned identically to how the regressors were tuned in the main text, as is discussed in Appendix C. The importances for each of the parameters is shown in Fig. D2. Comparing with Fig. 1, we can see that the order of the importances is basically unchanged, indicating neither of the two velocity dispersions is more important than the other in determining the Balmer decrement.

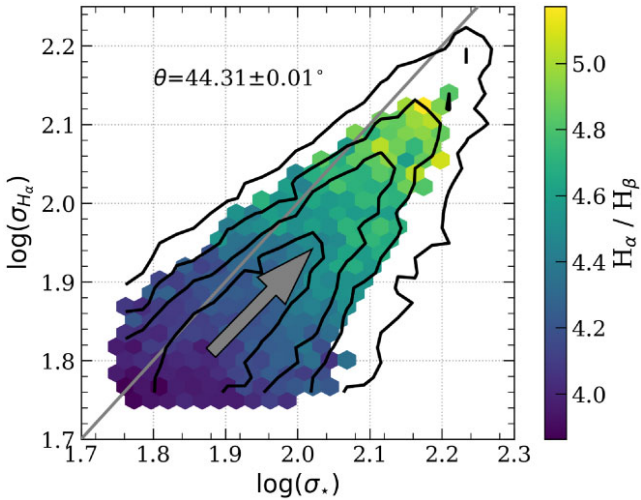

**Figure D1.** Nebular velocity dispersion as a function of the stellar velocity dispersion, colour-coded by the Balmer decrement for local galaxies (SDSS). The grey arrows denote the direction in which the Balmer decrement has the largest gradient, determined using the PCC coefficients, with its angle defined clockwise from the positive $y$-axis. The arrow angle indicates the dependence of the Balmer decrement on the two velocity dispersions is almost equal. The straight gray line has a gradient of unity to guide the eye. The galaxies are slightly biased to lower nebular velocity dispersion and higher stellar velocity dispersion, as expected from the literature (Crespo Gómez et al. 2021; Übler et al. 2022). The offset of the two velocity dispersion indicate they are not exactly interchangable. The black contours indicate the density of the galaxies in this parameter space, with the outermost contour containing 95 per cent of the galaxies.

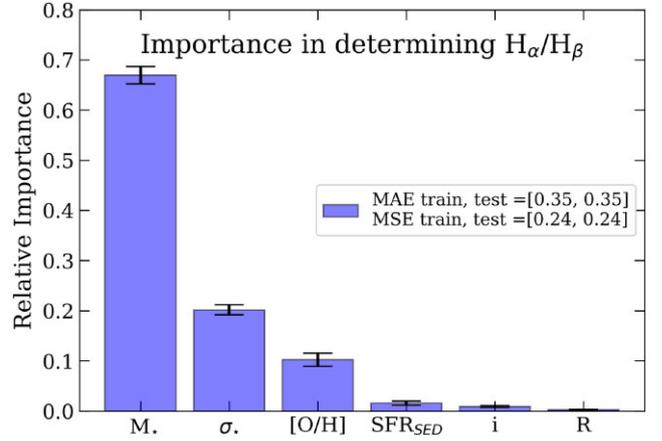

**Figure D2.** Plot showing the importances of the different galactic parameters in determining the Balmer decrement along with their errors for local galaxies (SDSS), considering $\sigma_\star$ instead of $\sigma_{H_\alpha}$. There is minimal difference between the importances determined here and those presented in Fig. 1, showing that each velocity dispersion is roughly as important as the other in determining the Balmer decrement. The difference between the average MAE and MSE of the train and test samples being so small implies no overfitting. Parameter R is the random variable and SFR$_{SED}$ is the SFR derived from SED fitting.

However, since stellar velocity dispersion measurements were not available for the galaxies at $z \sim 1-3$ used in this work, we decided to instead use the nebular velocity dispersion of all galaxies considered.

## APPENDIX E: DISPERSION MINIMIZATION

For a given $\alpha$ and $\delta$, the dispersion in the Balmer decrement of the local galaxies was determined by binning the galaxies in 0.15 dex bins of $\mu$ and taking the standard deviation of the Balmer decrement of the galaxies in each bin, if there were at least 25 galaxies in the bin. The total dispersion at the given $\alpha$ and $\delta$ was then the weighted average of those standard deviations, weighted by the number of galaxies in each bin of $\mu$. The latter requirement mitigated the effect of sparsely populated bins reducing the dispersion as the galaxies spread out in the parameter space at large rotations, creating artificially low dispersion at large rotations.

The approach taken in this work to identify the best fit in $\alpha$ and $\delta$ was iterative. We started by creating a grid in $\alpha$, calculated the dispersion at each grid point with $\delta = 0$ and identified the minimum. We then made a grid in $\delta$ and calculated the dispersion with the value of $\alpha$ at the minimum. This process was iterated until the minimum parameter values converged. Convergence was reached when the subsequent value of $\alpha$ and $\delta$ were equal within 16 dp to the preceding value.

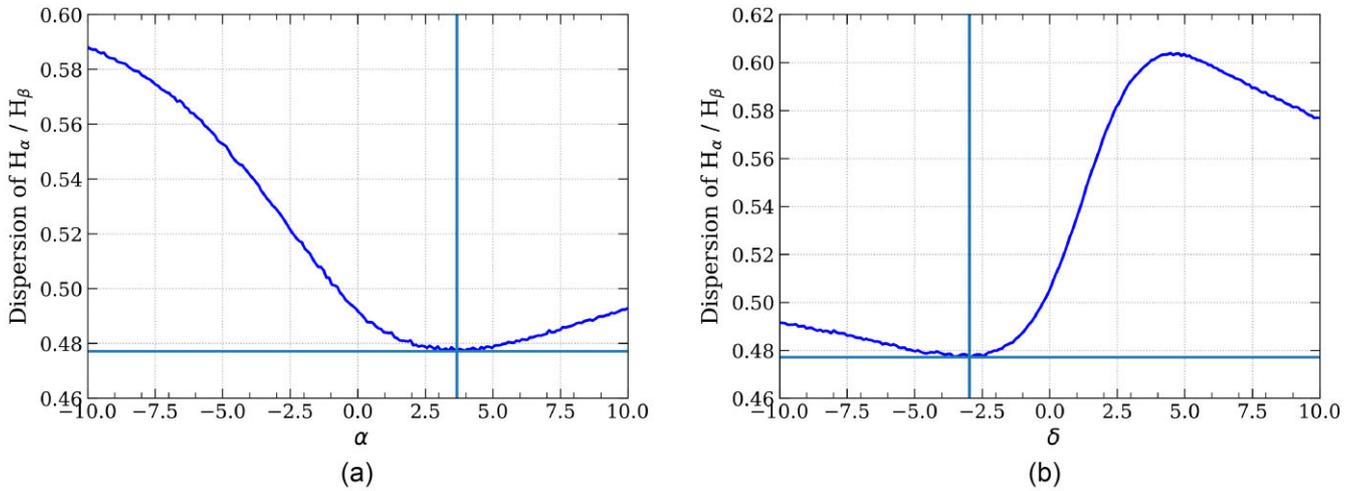

(a)　　　　　　(b)

**Figure E1.** Plots of the dispersion in the Balmer decrement of the local galaxies (SDSS) against the minimization parameter $\alpha$ (a) and $\delta$ (b) for the final iteration of the minimization analysis. The minimum of the dispersion curve and the corresponding value of the minimization parameter is shown by the horizontal and vertical lines, respectively. The minimization parameters are defined through $\mu = \log M_\star + \alpha[O/H] + \delta \log \sigma_{100}$ as in equation (7).

For each parameter, the grid had 200 points between −10 and 10. After three iterations, the minimization converged, and these final minimization curves for $\alpha$ and $\delta$ are shown in Fig. E1. The final values are $\alpha = 3.67$ and $\delta = 2.96$.

This paper has been typeset from a TEX/LATEX file prepared by the author.